\documentclass[11pt, preprint]{aastex}

\usepackage{natbib}
\usepackage{graphicx}
\usepackage{amsmath}
\usepackage{amssymb}
\usepackage{epstopdf}
\usepackage{ulem}
\usepackage{color}

\usepackage{overpic}
\usepackage{pict2e}

\newcommand{\eg}{{\it e.g.}}

\newcommand{\rsun}{R_\odot}
\newcommand{\bx}{{\bf r}}

\newcommand{\cc}{{\mathcal C}}

\newcommand{\khat}{\hat {\bf k}}

\newcommand{\bnabla}{{\boldsymbol \nabla}}
\newcommand{\curl}{{\boldsymbol \times}}
\newcommand{\bxi}{{\boldsymbol \xi}}

\newcommand{\bfm}{{\boldsymbol m}}
\newcommand{\bfv}{{\boldsymbol v}}

\begin{document}

\title{Seismic Sounding of Convection in the Sun}

\author{Shravan Hanasoge$^{1}$, Laurent Gizon$^{2,3}$ and Katepalli R. Sreenivasan$^{4}$
\affil{$^1$Tata Institute of Fundamental Research, Mumbai, India 400005}
\affil{$^2$Max-Planck-Institut f\"{u}r Sonnensystemforschung, Justus-von-Liebig-Weg 3, 37077 G\"ottingen, Germany}
\affil{$^3$Institut f\"{u}r Astrophysik, Georg-August-Universit\"{a}t G\"ottingen, 37077 G\"{o}ttingen, Germany}
\affil{$^4$New York University, NY, USA 10012}
}

\begin{abstract}
Our Sun, primarily composed of ionized hydrogen and helium, has a surface temperature of 5777~K and a radius $R_\odot \approx 696,000$ km. In the outer $R_\odot/3$, energy transport is accomplished primarily by convection. Using typical convective velocities $u\sim100\,\rm{m\,s^{-1}}$ and kinematic viscosities of  order $10^{-4}$ m$^{2}$s$^{-1}$, we obtain a Reynolds number $Re \sim 10^{14}$. Convection is thus turbulent, causing a vast range of scales to be excited. The Prandtl number, $Pr$, of the convecting fluid is very low, of order  $10^{-7}$\,--\,$10^{-4}$, so that the Rayleigh number ($\sim Re^2 Pr$) is on the order of $10^{21}\,-\,10^{24}$. Solar convection thus lies in extraordinary regime of dynamical parameters, highly untypical of fluid flows on Earth. Convective processes in the Sun drive global fluid circulations and magnetic fields, which in turn affect its visible outer layers (``solar activity") and, more broadly, the heliosphere (``space weather"). The precise determination of the depth of solar convection zone, departures from adiabaticity of the temperature gradient,  and the internal rotation rate as a function of latitude and depth  are among the seminal contributions of helioseismology towards understanding convection in the Sun. Contemporary helioseismology, which is focused on inferring the properties of three-dimensional convective features, suggests that transport velocities are substantially smaller than theoretical predictions. Furthermore,  helioseismology provides important constraints on the anisotropic Reynolds stresses  that control the global dynamics of the solar convection zone. This review discusses the state of our understanding of convection in the Sun, with a focus on helioseismic diagnostics. We present our considerations with the interests of fluid dynamicists in mind.
\end{abstract}



\keywords{Helioseismology, Convection, Non-Boussinesq, Dynamo}

\section{Introduction} 

Solar energy is powered by nuclear fusion in the Sun's core (inner 20\% by radius) through p-p chain fusion, whereby ionized hydrogen is converted to helium. Other by-products of fusion include high-energy photons and neutrinos. Density and temperature are very high in the core, on the order of 150 g/cm$^3$ and $15.7$~million K, respectively. High-energy photons released in fusion reactions are absorbed in a few millimeters of solar plasma and then re-emitted in random directions and at lower energies ({\it free-free scattering}). Photons gradually diffuse outwards from the core to the Sun's surface; random walk estimates suggest a transit time in excess of 100,000 years  (while neutrinos take only a few seconds to reach the surface). Since the temperature gradient is sufficiently small below about $0.7 R_\odot$, radiative transfer by photons is the primary means of energy transport from the core. At the base of the convection region, i.e. $r\approx0.7 R_\odot$, the temperature is approximately 2 million K. Radiation becomes less efficient further outwards because of the increase in the opacity of the plasma, and much of the energy transport occurs by thermal convection. The radiative interior and the convective zone are separated by a thin layer \citep[$\approx 0.05 R_\odot$, \eg][]{elliott99, monteiro11} where the stratification changes rapidly from convective stability to marginal instability. This region also shows a relatively sharp change between the solid-body rotation of the radiative interior and the differential rotation of the exterior convection zone; it is termed the tachocline \citep{spiegel92}.

Convective motions are thought to drive the dynamics and activity of the Sun. On small scales, convection directly generates and sustains magnetic fields \citep[the Sun's small-scale dynamo field; \eg][]{voegler2005}. On larger scales, the turbulent convection in the rotating Sun produces Reynolds stresses which drive large-scale (global) motions. These large-scale motions amplify the global magnetic field whose manifestations at the solar surface are sunspots of magnetic complexes, which in turn are stressed by convective motions, producing flares and space weather. Magnetic fields act on long timescales to mediate angular momentum loss from the Sun as a whole (magnetic braking). Convection also mixes plasma inside the Sun; this type of mixing can significantly affect the lifetimes of main-sequence stars (that is, stars that are fusing hydrogen atoms to form helium atoms in their cores). For these reasons it is important to understand convection inside stars, particularly the Sun.

Convective motions in the Sun are more complex than in laboratory experiments of Rayleigh-B\'enard convection, in particular because of stratification. For instance, the density contrast between the base and the top of the solar convection zone is about $10^5$. Furthermore, the energy transport by convection is very efficient in the Sun, in that only very small departures from the adiabatic temperature gradient are required. These motions are also difficult to model in faithful detail: the relevant parameters in the Sun are extreme, with the highly turbulent flows involving spatial and temporal scales spanning many orders in magnitude. For reviews on solar convection, see \eg\, \citet{miesch05}, \citet{miesch09}, \citet{rieutord10}, and \citet{hanasoge14}.

In this context observational tests of our understanding are essential, and we shall focus in this review on the contributions made by helioseismology. The accurate inference of the depth of the convection zone, the measurement of the departure from adiabaticity as a function of depth, and the determination of solar interior rotation and meridional circulation (at least in the surface layers) are all important contributions of helioseismology \citep{jcd02,gizon05}. The current focus is on characterizing the three-dimensional (3-D) properties of convective flows and magnetism in the solar interior using techniques of local helioseismology \citep{gizon2010}.

\subsection{Comparison with laboratory convection}

Standard experimental work on convection is carried out in the Rayleigh-B\'enard apparatus. Nominally, a shallow fluid layer contained in a cylinder with non-conducting side walls and conducting bottom and top walls is heated from below and cooled from above, so that the entire heat from the bottom wall goes through the fluid to the top wall. An important question concerns the amount of heat that is transported across the fluid for a given temperature difference maintained between bottom and top walls. Heat is transported purely by molecular conduction if the temperature difference is small (in a sense to be clarified), but the fluid begins to move and enhance the heat transport thereafter. A non-dimensional measure of the heat transport is the Nusselt number, which is the ratio of the actual heat transport to that possible by conduction alone. The buoyancy effect produced by the temperature difference is balanced by the smearing effects of diffusion and viscosity, the ratio of the former to the latter being called the Rayleigh number; the larger the Rayleigh number the larger is the effective temperature difference. One goal of convection studies is to discover how the Nusselt number scales with the Rayleigh number. Much is known about this flow \citep{Castaing1989, niemelasreenivasan2006, ahlers, chilla} though the nature of the heat transport law for asymptotically large Rayleigh numbers is still a topic of lively discussion \citep{Siggia1994, NS2003, NS2010, Ahlers2012, urban, Stevens2013, Shishkina2015}.

It is also possible to establish a flow in which one fixes the rate of heat transport at the bottom and top walls, instead of their temperatures. It is this latter condition that might apply to the Sun and other stars. The convection boundaries are evidently not solid but steep changes in the adiabatic temperature gradient that occur at the base of convection zone significantly restrict free movement of plasma across this layer. There can be small-scale penetrative convection; indeed, it is thought to play a critical role in stellar rotation spin down and, therefore, measuring the depth of penetration is an active area of research in helio- and astero-seismology \citep[\eg][]{monteiro11}. Nevertheless, all large scale motions are restricted at the base resulting in a plausible analogy with a solid surface. Likewise, at the top of the convection region, steep gradients in temperature and density caused within the near-surface convection region endow it with boundary-like properties. 

Thermal energy produced in the core and transmitted at this radiative-convective interface provides the constant heat flux boundary condition at the bottom of convection zone. However, the variation of thermodynamic properties and the gas composition between the top and bottom of the convection zone renders stellar convection highly non-Boussinesq. We recall that only small deviations from the large adiabatic gradients of temperature and density matter for convection. Yet the Rayleigh numbers are vast and motions inside the convection zone are turbulent.

It is useful to note that the Prandtl number of the fluid is very small everywhere in the convective flow regime of the Sun. Even fluids like mercury and liquid sodium, sometimes used in the laboratory, possess Prandtl numbers which are several orders of magnitude larger than solar plasma. This could have important consequences, but the role of the Prandtl number at very high Rayleigh numbers is not well understood. For small Prandtl-number fluids, no significant heat transfer enhancement is seen in laboratory measurements \citep{Ecke2014, ahlers} and direct numerical simulations \citep{Horn2015,Stevens2010b, Oresta2007} of rotating turbulent Rayleigh-B\'enard convection. Ekman vortices develop irregularly and are shorter than in the case of large Prandtl-number fluids. The thermal diffusivity, larger than the kinematic viscosity, reduces the effectiveness of Ekman pumping \citep{Stevens2010b, Horn2015}. For all Prandtl numbers, \citet{Horn2015} identify the different regimes of rotating Rayleigh-B\'enard convection by comparing the poloidal and toroidal energies of the flow field. This identification was found to be robust to changes in the aspect ratio and geometry of the container.

While the entire region of the Sun beyond $r\approx0.7\rsun$ is convective, it may be conceptually useful here to consider the outer $0.3\rsun$ as two separate regions. ``Near-surface convection" is a thermal boundary layer with a thickness on the order of 300 km, where radiative opacities change rapidly and the plasma transitions from optically thick to thin, i.e., it becomes more transparent. The rest of the convection region, forming the bulk, is termed the ``deep convection'' region. These two regions of the convection zone are likely connected by the cool descending plumes associated with overturning surface convection, seeded at the photosphere. The fate of these plumes, i.e. their overall stability during descent sets the degree of mixing and thus the dynamics of global convection \citep[\eg][]{spruit97,nordlund09}.

\subsection{Near-surface convection} 
Near-surface convection, marked by the presence of convective cells of characteristic size 1000-2000 km called granules, can be observed by high-spatial-resolution optical imaging devices 
\citep[see][and references therein]{nordlund09}.   
The centres of granules are hotter than the edges, implying that granulation is a form of overturning convection, where hot upward moving plasma radiates away the thermal energy, and channels between granules (intergranular lanes) provide conduits for the cool fluid to reenter the solar interior. Granules are the primary vehicles of thermal transport in the near-surface layers of the Sun, carrying nearly the entire solar luminosity outwards \citep{nordlund09}. Granules have been the subject of careful observation \citep[\eg][]{schrijver97} and properties such as their size and intensity-contrast are well characterized. Comprehensive numerical simulations of surface convection \citep{stein00,voegler2005} are successful in reproducing these properties. 
	
\begin{figure}
\centering
\includegraphics[width=\linewidth]{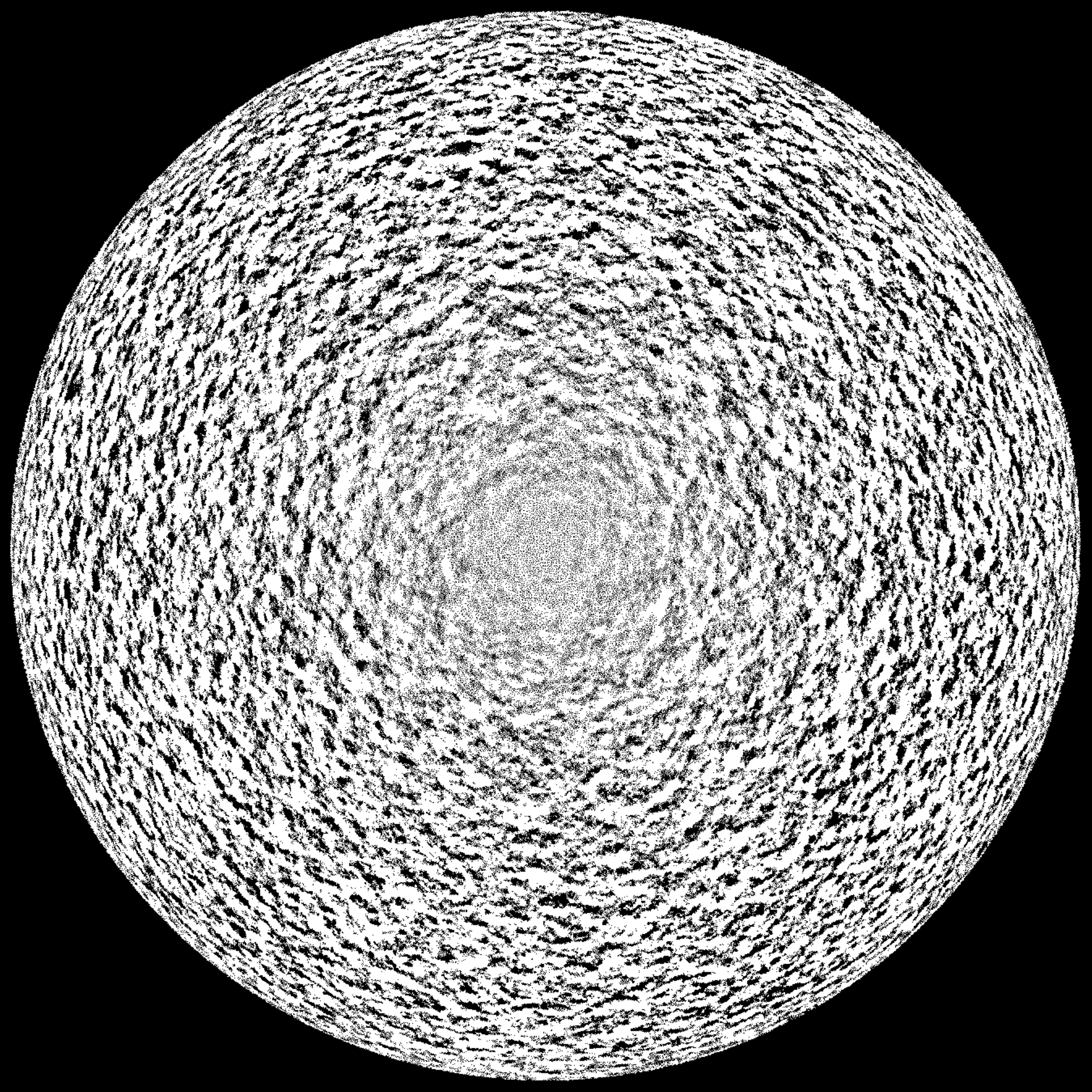}
\caption{Line-of-sight projected velocity field measured at the photosphere of the Sun by the Michelson Doppler Imager. Average rotation has been subtracted. The structure visible in this image is supergranulation, a flow system of size 35000-km in which horizontal velocities are about 300~m/s and radial velocities are $\sim30$~m/s at the surface. Courtesy M. Rieutord, see \citet{rieutord10}.}
\label{supergran}
\end{figure}
 
One challenge of the simulations concerns  {\it supergranulation}, a distinct pattern of horizontal flows (root-mean-square velocity on the order of 300~m/s) with a characteristic size of 35\,000 km and a lifetime of 1\,--\,2 days, which is easily seen away from the disk center in direct Doppler images \citep[see Figure~\ref{supergran};][]{rieutord2010, Hathaway2000}. Although supergranulation was discovered by \citet{Hart1954} more than sixty years ago, its origin and subsurface properties still pose major challenges. It has been suggested that it may correspond to a preferred scale of deep convection \citep{leighton62} or to a nonlinear dynamical interaction of granules \citep{rast03,supergranulation10}. The supergranulation pattern propagates faster than the solar plasma and has been cited as an example of traveling-wave convection \citep{gizon03}.

\subsection{Deep convection}
Convection in the interior is a difficult problem to address because it is both optically inaccessible and the governing physics lies in a parameter regime that is beyond laboratory studies or numerical simulations. The basic issue of whether deep convection is controlled by heating from below or by the radiative cooling from above or by a combination of the two is not clear \citep{spruit97}. Owing to steep stratification (nonetheless far smaller than that in the near-surface convection region), fluid properties such as the kinematic viscosity and thermal and magnetic diffusivities at the base and top of the convection zone, and thus the time and length scales in those regions, are dramatically different. Therefore if heating and cooling were to be decoupled, i.e. if mixing of descending plumes into the surrounding medium were weak, it is conceivable that the problem would separate into more tractable sub-problems. Whereas, if the coupling were strong, the physics of the convection zone would be altogether stiff since both small and large scales would play crucial roles. The physics of the coupling of convection at the surface thermal boundary layer with deeper layers is thus less easy to understand than in standard Boussinesq convection.  

\begin{figure}
\centering
\includegraphics[width=0.7\linewidth, trim=60 360 60 80,clip]{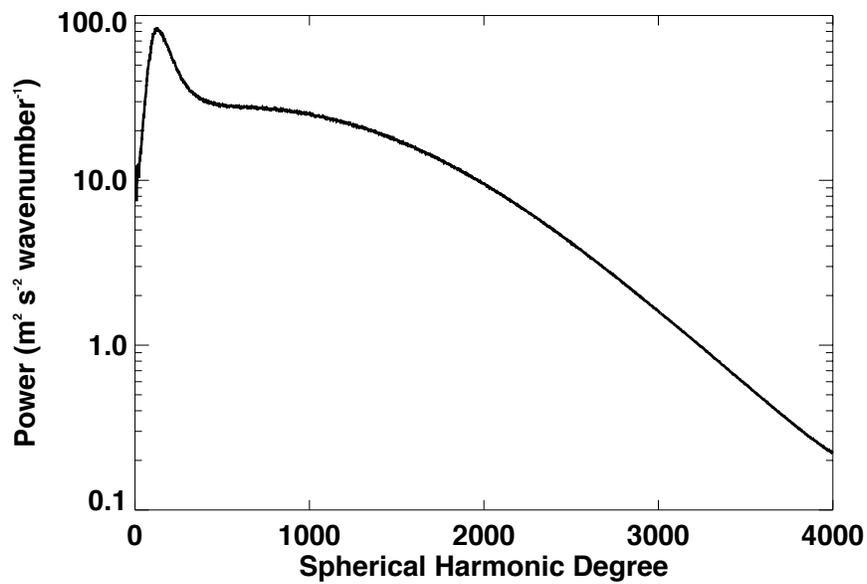}
\caption{Power spectrum of surface convection as a function of spherical-harmonic degree, $\ell$, as measured in line-of-sight Doppler velocity by the Helioseismic and Magnetic Imager.  Granulation energy  is broadly spread up to $\ell \sim 4000$. A distinct peak corresponding to supergranulation is seen around $\ell\approx120$. At low degrees, convective velocities fall sharply  with decreasing $\ell$. 
Courtesy of David Hathaway.}
\label{spectrum}
\end{figure}

 Numerical simulations typically show large-scale overturning convective cells, coherent on scales much greater than supergranulation. In other words, the surface convective power spectrum obtained from numerical calculations suggest that the largest scales are most energetic, in contrast to that on the surface. Simulations show elongated ``banana cells'' at low latitudes \citep[\eg][]{miesch_etal_08}, which have not (yet) been detected on the Sun. Using measurements of the photospheric flow field, \citet{hathaway2013} measured large-scale fluid motions (giant cells), with velocities of the order of 8\,-\,20 m/s, smaller by at least an order in magnitude than those associated with intermediate and other smaller scale convective structures (also see Figure~\ref{spectrum}). Giant cells are seen at all latitudes, with a seeming preference for high ones.

One might speculate that the limited parameter regime of numerical calculations can only lead to standard turbulent mixing and conventional overturning convection. The fundamental question is whether, in the dramatically different regime prevalent in stellar interiors (that of the Sun in particular), the highly asymmetric up and downflows are well mixed at all (also see section~\ref{weakconvection}). Differential rotation in the convective envelope provides a stringent test of simulations, and prior efforts have failed to reproduce it accurately from first principles \citep[\eg][]{balbus12,miesch12}. Another constraint on the quality of the computed solutions is the prediction of meridional circulation, although uncertainties in our knowledge of meridional flow limit its potency as a constraint \citep[\eg][]{gizon2010}. In spite of inadequate resolutions, numerical simulations provide the best and detailed estimates of the properties of convection and dynamo action. 
 
The base of the convection zone contains a layer where the temperature gradient sharply transitions to strong stability.
The motion of fluid is thus restricted and small-scale plumes penetrate into the overstable radiative interior. Convective penetration affects the mixing of chemical elements in the interior. It also excites internal gravity waves which in turn contribute to the transport of angular momentum and to the secular spin down of the Sun \citep{mathis13}. The physics of convective overshoot is complicated by the presence of magnetism and the strong coupling between enormously different timescales in the convection zone and radiative interior. Appreciating the dynamics of this layer is actively pursued through the use of phenomenological models \citep[\eg][]{zahn91}, numerical simulations \citep[\eg][]{rogers05,brun11}, and seismic analyses \citep[\eg][]{monteiro11}.

\section{Models of convection}
\subsection{Mixing-length theory}
Since convection plays a crucial role in stellar evolution (by the mixing of elements), simplified prescriptions for its properties are of considerable importance. A widely used model is the mixing-length theory (MLT) of convection, originally due to \citet{prandtl25} for hydrodynamic turbulence, extended to stellar physics by \citet{bohm58}. Subsequently, MLT has been used to model the solar convection zone \citep{spruit74,gough77} and is now widely used in stellar evolution codes. MLT is based on the simple phenomenology of parcels of fluid transporting thermal energy over a specified length scale and subsequently immediately undergoing complete mixing. On the other hand, turbulence and convection are multi-scale phenomena. This crucial aspect of MLT, namely a single-scale description, does not handle the physics adequately \citep[\eg][]{cox_2004}. Further, MLT does not model strong compressibility and interaction with radiation associated with near-surface layers, making it inaccurate there. Nevertheless, it provides a basic recipe for incorporating convection into stellar models and is therefore of great utility. More sophisticated modeling is required to address multi-scale dynamics and radiative transfer \citep[for instance, see][for more detailed models of convection]{canuto98}.

\begin{figure}
\centering
\includegraphics[width=0.8\linewidth]{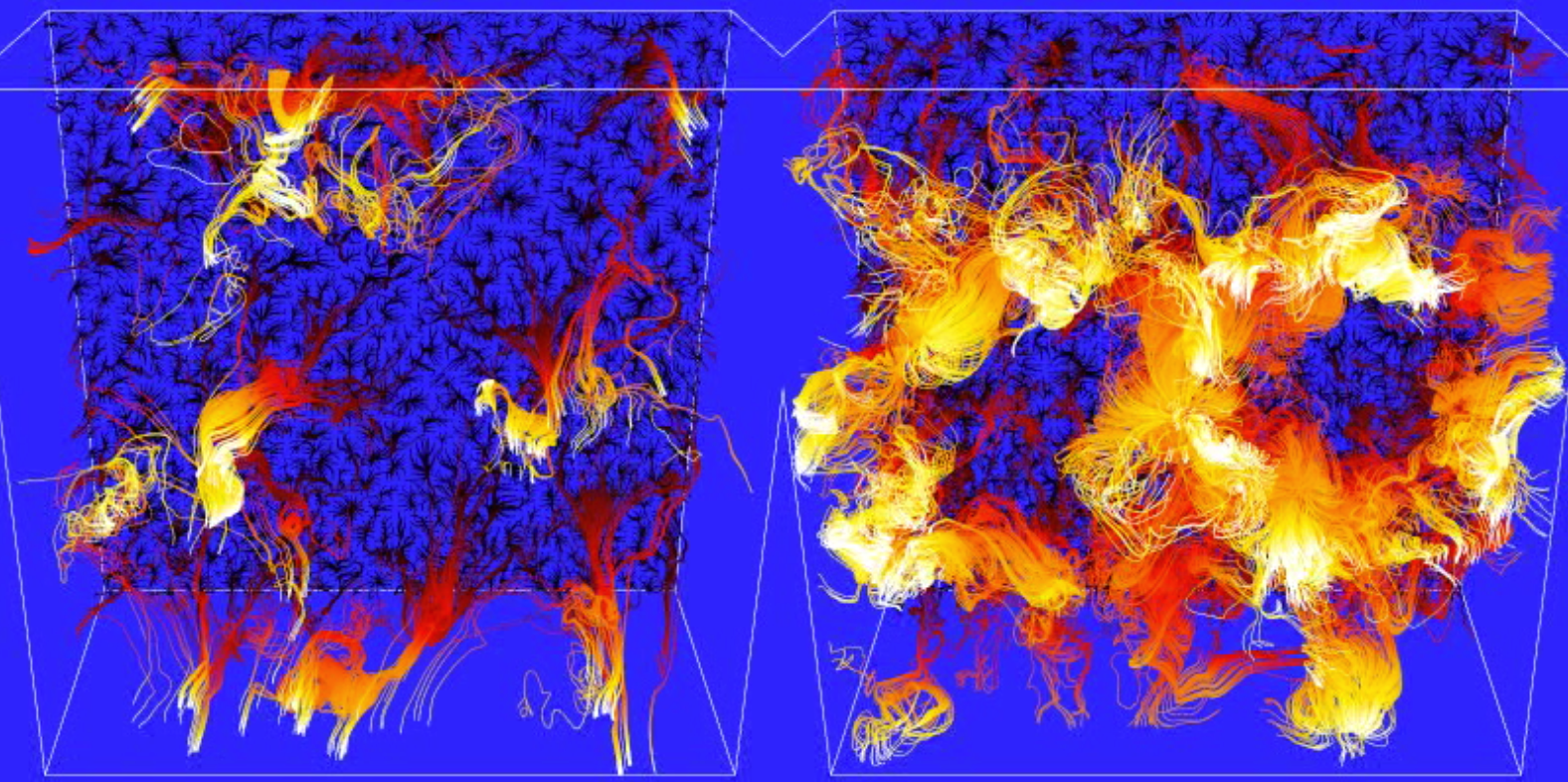}
\caption{Streamlines from a fully-compressible numerical simulation of near-surface convection
\citep[Stagger code;][]{nordlund09}.
The computational domain (a box 48~Mm on the side and 20~Mm deep) is seen from the bottom.
On the left are the streamlines for up flows measured in the photosphere, on the right for downflows.
The color along the streamlines indicates depth (white for deep, dark red for the surface).  
The downflows that reach the bottom of the box delineate the borders of supergranules.
The upflows that originate from the deep layers are found at the center of supergranules.
Courtesy of Robert Stein.
}
\label{streamlines}
\end{figure}
\begin{figure}
\centering
\includegraphics[width=0.8\linewidth]{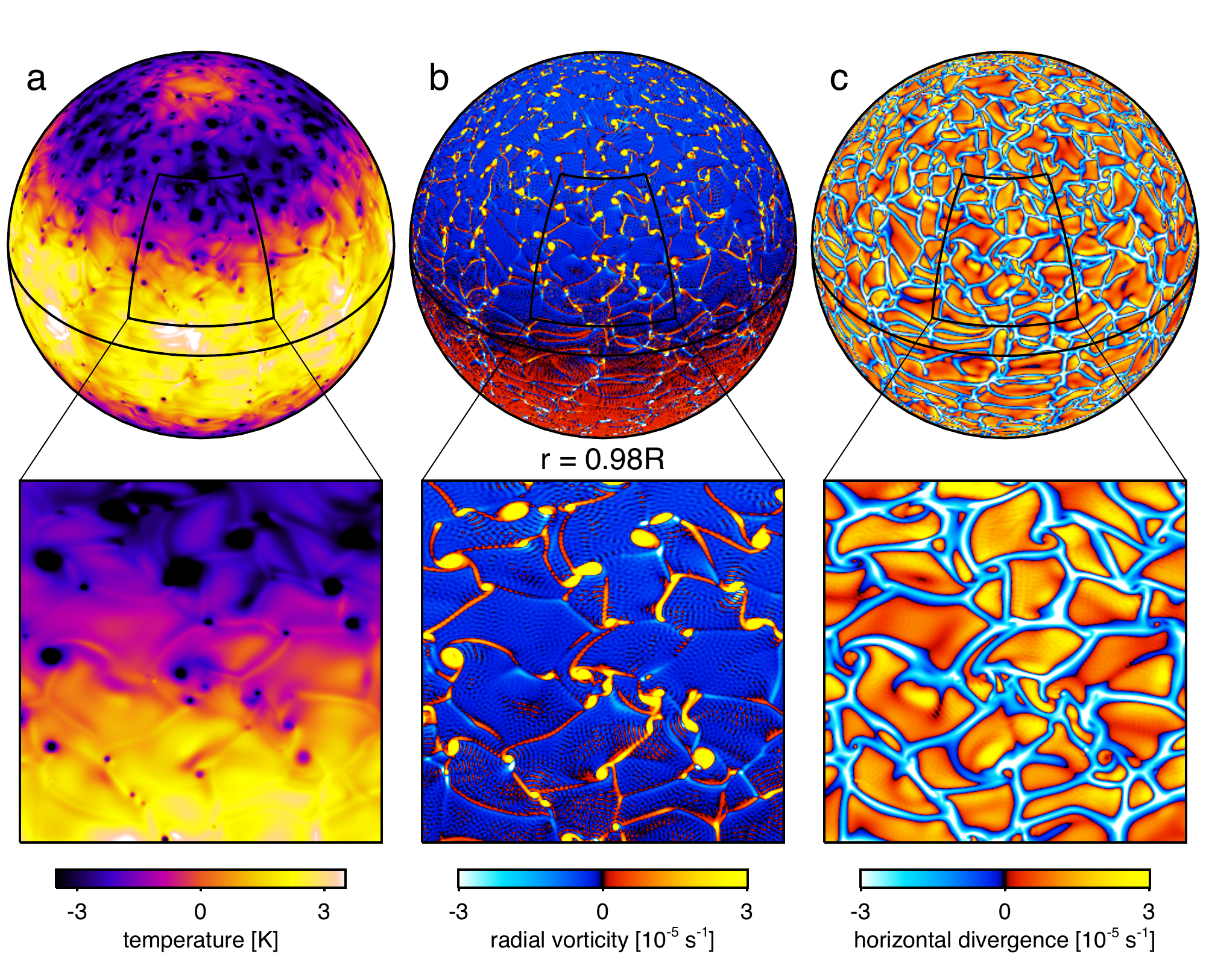}
\caption{Temperature, radial vorticity, and horizontal divergence near the top ($r=0.98 R_\odot$)
of a simulation of  deep solar convection in a spherical shell (ASH simulation).
The equator is indicated with a solid line. Magnified areas shown in the lower panels 
are square regions of size $45^\circ$. From \citet{miesch2008}.
}
\label{deepconv}
\end{figure}

\subsection{Numerical simulations}
The surface layers of the Sun exhibit dynamic interplay between small-scale magnetism and granulation. Observations of granulation show a flow field that has distinct, seemingly well formed convective cells of typical size 1000\,--\,2000 km, with a lifetime of about 10 minutes. Fibril-like magnetic fields are swept to the intergranular lanes by horizontal flow fields \citep[\eg][]{graham10}. Granulation is also the primary source of acoustic oscillations observed on Sun's surface \citep[\eg][]{goldreich77,nordlund09}. Simulating near-surface convection requires modelling radiative transfer and compressible hydrodynamics in a strongly stratified medium. 
Given these complexities, the accurate reproduction of the observables (such as granule size distribution, flow magnitude, and line formation properties) by simulations is remarkable \citep[\eg]{stein00,voegler2005,rempel09,nordlund09}; this is attributable perhaps to the steep density stratification (scale height of $\sim$ 200 km), which causes the flow field to expand rapidly with height, smoothing small-scale fluctuations and limiting the degree of interaction between the large and small scales. 

The computation of deep convection is a much more challenging problem, requiring one to incorporate a vast range of spatial and temporal scales, inaccessible with current computing power. Nonetheless, a number of codes to simulate global spherical convection are in use \citep{miesch_etal_08,charbonneau10,kapyla1,kapyla2,hotta14}. Most of these methods invoke the anelastic approximation \citep{gough69}, a low-Mach-number limit of the Navier-Stokes equation for stratified media in which acoustic waves are filtered out. Removing acoustic waves substantially relaxes the Courant restriction on the time step, thereby resulting in computational savings. Alternately, \citet{hotta14} have developed a technique in which they artificially reduce the sound speed, allowing weak compressibility to be incorporated (in contrast to the anelastic limit).

\section{Helioseismology of solar convection}
Helioseismology provides an independent and powerful means to constrain properties of interior convection. The Sun and its main-sequence cousins support a spectrum of oscillations, continuously excited by vigorous turbulent motion. The descending plumes in the intergranular lanes accelerate as they fall inward and are thought to generate shocks, thereby channeling their energy into acoustic radiation \citep[\eg][]{leibacher71}. While the generation of acoustic waves is a nonlinear process, the subsequent wave propagation is well described by linear theory owing to the fact that solar acoustic mode amplitudes are substantially smaller than sound speed. Solar oscillations have resulted in accurate inferences of the structure of the Sun \citep{jcd}, its internal rotation \citep[\eg][]{schou98}, and other large-scale flow features such as meridional circulation \citep{giles97} and toroidal oscillations \citep{howe00}.

Seismic inference involves measuring oscillation frequencies or wave travel times, and building models of a medium in which the predicted properties of wave propagation would match observations. Solar oscillations were discovered by \citet{leighton62}, who speculated that these waves could be used to infer the properties of the surface layers. Subsequent measurements showed that the Sun exhibits acoustic waves that are trapped in the radial direction \citep[for a review, see \eg][]{jcd02}, which could be used to probe the structure and dynamics of its deep interior. Contemporary helioseismology deals with the study of non-axisymmetric, three-dimensional perturbations due to magnetic fields and convective flow systems \citep[\eg][]{gizon2010}. In subsequent discussions, we invoke the term ``horizontal''  which refers to the latitude-longitude directions and ``radial" to the spherical radial direction. In many instances we perform the seismology of small regions, small in comparison to the solar radius. The formal interpretation of these measurements is expressed in plane-parallel Cartesian geometry, and the term``radial" is replaced by ``vertical". Two robust quantities that are typically inferred from seismology are the horizontal flow divergence and radial component of flow vorticity, i.e., $\bnabla_h\cdot{\bfv}$ and $\bnabla_h\curl{\bfv}$, where $\bnabla_h$ is the horizontal component of the gradient operator and $\bfv$ is the flow velocity \citep[see][for a description of how these measurements are made]{Langfellner2014}.   

\subsection{Seismic measurements}
The Helioseismic and Magnetic Imager \citep[HMI;][]{hmi}, an instrument onboard the Solar Dynamics Observatory launched by NASA in 2011, provides an important source of helioseismic observations. Among other measurements, HMI takes $4096\times4096$-pixel images of the line-of-sight velocity at the solar surface every 45~seconds (Doppler shifts of the Fe-I absorption line). Ground-based instruments such as the Global Oscillation Network Group (GONG) also provide streams of observations. Seismic waves are continually excited by turbulent convection in the Sun and consequently the wavefield is stochastic. Following  \citet{duvall}, we measure the two-point correlation function $\cc$ of the wavefield $\phi(\bx,t)$ from finitely long observational records,
\begin{equation}
\cc_{\alpha\beta}(t) = \frac{1}{T}\int_0^T \phi(\bx_\alpha, t')\,\phi(\bx_\beta,t'+t)\, dt' ,
\label{crossc}
\end{equation}
where $T$ is the duration of the observation, $t$ is the correlation time lag, and $\bx_\alpha$ and $\bx_\beta$ are two locations on the surface. The Sun is continuously evolving with time and so averages over some finite temporal extent $T$ imply that the dynamics of interest is quasi-static over $T$. For each pair of points, the travel time for waves propagating from $\bx_\alpha$ to $\bx_\beta$, denoted by $\tau_{\alpha\beta}$, can be extracted from the $t>0$ part of the cross-correlation function \citep{gizon02,gizon05}. In turn, travel times may be formally interpreted \citep{gizon02,hanasoge11} in terms of products of Green's functions of operator~(\ref{goveq}) and an inverse problem can be posed. The travel time differences $\delta\tau_{\alpha\beta} = \tau_{\alpha\beta}-\tau_{\beta\alpha}$ are particularly sensitive to flows (as opposed to isotropic wave-speed perturbations) and are thus used as input to the inverse problem.

\subsection{Seismic modeling}
Wave propagation in the Sun can be very effectively treated as a linear problem owing to their small amplitudes in comparison to the sound speed \citep[wave amplitudes are on the order of cm/s compared to km/s sound speeds;][]{jcd02}. The displacement vector $\bxi$ associated with small-amplitude  waves satisfies \citep[\eg][]{ostriker67}
\begin{eqnarray}
\rho (\partial_t - \bfv\cdot\nabla)^2 \bxi - \bnabla(\rho c^2\bnabla\cdot\bxi + \bxi\cdot\bnabla p) - {\bf g}\,\bnabla\cdot(\rho\bxi)  = \bf S    , \label{goveq}
\end{eqnarray}    
where $\rho$, $p$ and $c$ are the  background density, pressure, and sound speed at position $\bx$ and time $t$    and  ${\bf g} = - g(r)\, {\hat{\bf r}}$ is the acceleration due to gravity.  In equation~(\ref{goveq}), the first two terms model sound-wave propagation in a stratified medium and the third term is the effect of buoyancy. Advection by a background flow 
${\bfv}$ is included. In order to facilitate seismic analyses, we assume that the background properties such as density, flows, pressure and sound speed vary on time scales much longer than the acoustic wave period and the observation duration. The source of excitation $\bf S$ represents random forcing by small-scale turbulence (granulation).

Seismology is a technique to infer the wave speed distribution in the medium. In a conducting plasma, there are two basic types of wave speeds: an isotropic sound speed and magnetic Alfv\'{e}n velocity \citep[e.g.,][]{hanasoge12_mag}.  Additionally, flows in the medium advect acoustic waves in a manner that breaks the directional symmetry of propagation. Waves travel faster along the flow than against the flow. For instance, if the travel time between two points were to be $\tau$, the travel-time perturbation $\delta\tau$ introduced by a subsonic flow ${\bfv}$ is approximately given by $\delta\tau \approx - \tau\, {\bfv}\cdot\khat/c$, where $\khat$ is the unit wavevector. Thus a wave propagating along the direction of the flow will speed up, thereby reducing the travel time, and vice versa.

Formally, seismic inference requires solving an inverse problem to estimate the coefficients of the partial differential equation~(\ref{goveq}), such that the difference between observation and prediction is minimized in some prescribed measure (such as the $L_2$-norm difference). Given a model of the medium ${\bfm}=(\rho, c, \bfv \cdots)$, we establish a cost functional $\chi$ in terms of the travel times $\tau$, and thus $\bxi$, which in turn is affected by the medium according to equation~(\ref{goveq}). {An example of a travel-time misfit functional is
\begin{equation}
\chi = \frac{1}{2}\sum_{\alpha,\beta} (\tau_{\alpha\beta} - \tau_{\alpha\beta}^{\rm obs})^2,
\end{equation}
where the superscript 'obs'  denotes observations and $\tau_{\alpha\beta}$ the predicted travel time between points $\bx_\alpha$ and $\bx_\beta$. In this case, the goal would be to minimise the difference, measured in an $L_2$ sense, between the observed and predicted travel times.
We consider the variations of $\chi  = \chi(\bfm)$,
\begin{equation}
\delta\chi(\bfm) = \chi(\bfm + \delta\bfm) - \chi(\bfm) = \frac{\partial\chi}{\partial\bfm} \cdot\delta\bfm + O(||\delta\bfm||^2).
\end{equation}
To ensure convergence to the solution, i.e. $\delta\chi < 0$, we choose $\delta\bfm$ based on the negative gradient of $\chi$ with respect to $\bfm$, and use techniques such as steepest descent or nonlinear conjugate gradient to obtain improved estimates of $\bfm$. Methods of optimization constrained by partial differential equations \citep[also widely used in aerodynamic shape optimization, \eg][]{jameson88} can be used to arrive at a best-fit seismic model of the medium \citep{hanasoge11}. In this process it is particularly important to take account of the correct noise matrix, which can be directly estimated from observations or using a model \citep{gizon_04,fournier14}. More traditional approaches involve solving a linear inverse problem (the first iteration in the above problem); such approaches rely on careful choices for regularization terms \citep[\eg][]{Svanda2011,jason12}.

\subsection{Weak convection on large scales}\label{weakconvection}
An important first step in understanding convective flows is to provide estimates of their kinetic energy. To address this problem, \citet{hanasoge12_conv} analyzed HMI observations of the seismic wavefield using techniques of time-distance helioseismology \citep{duvall}. These raw data were downsampled, cross correlations were measured according to equation~(\ref{crossc}), spatially averaged, and travel times measured. The interested reader may refer to \eg\,\,\citet{duvall03} and \citet{hanasoge10} for details; the study by \citet{hanasoge12_conv} inferred velocities at depth for spherical harmonic degrees $\ell < 60$ to be smaller than those predicted by MLT and numerical simulations \citep{miesch08, charbonneau10, kapyla1} by one to two orders of magnitude. They are also below the surface (Doppler) velocities \citep[\eg][]{Hathaway2000, Rieutord2008}, which are themselves below the simulated values. These results are shown in Figure~\ref{constraints}. Also shown in the figure are measurements of giant cell velocities on the order of 8\,--\,20 m/s inferred from supergranulation (SG) tracking by \citet{hathaway2013}. The corresponding kinetic energy is about an order in magnitude larger than seismic constraints at depth $r/R_\odot = 0.96$.

\begin{figure}
\centering
\includegraphics[width=0.7\linewidth]{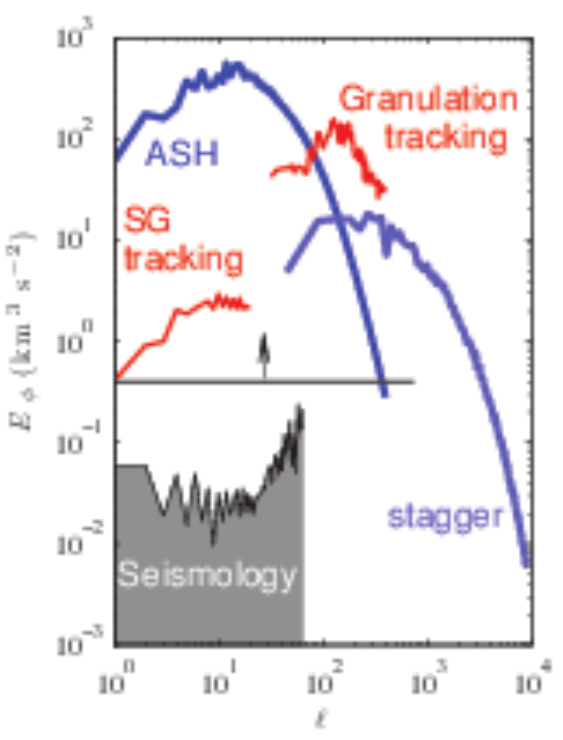}
\caption{Upper bounds on kinetic energy $E_\phi$ of longitudinal velocities $v_\phi$  versus  spherical-harmonic degree, $\ell$.  We define $E_\phi$ at radius $r$ such that $\langle v_\phi^2 \rangle /2  = \sum_{\ell\geq0} E_\phi(\ell)/r$, where the expectation value is approximated by a horizontal average. 
The grey region shows the helioseismology bound, based on $T=96$-hour samples of HMI observations \citep{hanasoge12_conv}. These constraints  are several orders in magnitude smaller than numerical simulations of global convection \citep[Anelastic Spherical Harmonic simulation (ASH);][]{miesch08}, suggesting that our current modeling of large-scale convection in the Sun is incomplete. 
The various curves denote convective energy spectra at different depths: seismology corresponds to $r/R_\odot = 0.96$, ASH to $r/R_\odot = 0.97$, and Stagger simulations to $r/R_\odot = 0.98$.
The horizontal black line is a theoretical lower bound based on global dynamics arguments \citet{miesch12}, assuming mode equipartition over $\ell < 750$.
The red curves are surface spectra based on HMI observations of granulation and supergranulation (SG) tracking. The SG tracking spectrum is based on data from \cite{hathaway2013}, courtesy of David Hathaway. Adapted from \citet{gizon12}. 
}
\label{constraints}
\end{figure}

Taken at face value, the seismic and giant-cell constraints of Figure~\ref{constraints} suggest that our understanding of thermal transport in the convection zone is incomplete. Because of the limited resolution of simulations and the weaker density gradients in them, the treatment of the highly asymmetric up and down flows is possibly not in the correct regime. Thermal transport of a solar luminosity can still be explained by decoupling the up and down flows and assuming that descending intergranular plumes make it all the way to the base of the convection zone \citep{spruit97}. 

A related challenge lies in explaining the convective power fall-off at low spherical-harmonic degrees. Using phenomenological models, \citet{lord14} attempted to derive a set of conditions that would result in this fall off. In line with growing evidence supporting weak convective motions on large scales, they conclude: ``we found that reducing the convective transport role of large-scale modes (by employing an artificial energy flux at all depths below 10 Mm which reduces the deep rms velocities by a factor of $\sim2.5$) can significantly improve the match between the coherent structure tracking spectra wof the simulations and observations. These separate lines of evidence all suggest that the Sun transports energy through the convection zone while maintaining very low amplitude large-scale motions. Something is missing from our current theoretical understanding of solar convection below $\sim10$ Mm''. Of particular relevance to models of convection are the matching of the amplitude of convective motions at all scales at the surface and accurately matching the ratios of power spectra at different depths. Both of these factors are strong tests of the mechanism of convection; the constraints in Figure~\ref{constraints} suggest that the amplitude and depth-scaling of the convective spectrum are in question. 
More work is needed to have greater confidence in the helioseismic inferences on deep convective velocities. In this regard, several ideas for improvement have been identified by \citet{gizon12}. In particular, the treatment of the interaction of solar acoustic waves with turbulence is not fully accounted for in current analyses \citep[\eg][]{hanasoge_bal13}.

The ratio of the horizontal dimension to the radial dimension of the convection region is the aspect ratio. The effective aspect ratio of the convective zone in the Sun is large (of order 10) but not in most high-Rayleigh-number laboratory studies (of order 1). Nevertheless, some large-scale structural features of convection are shared between laboratory experiments and observations. Higher aspect ratios support large-scale motion that extends over the entire extent of the apparatus, but it is relatively weak; see \cite{NS}. 

\subsection{Flow divergence and vorticity}

\begin{figure}
\centering
\includegraphics[width=\linewidth]{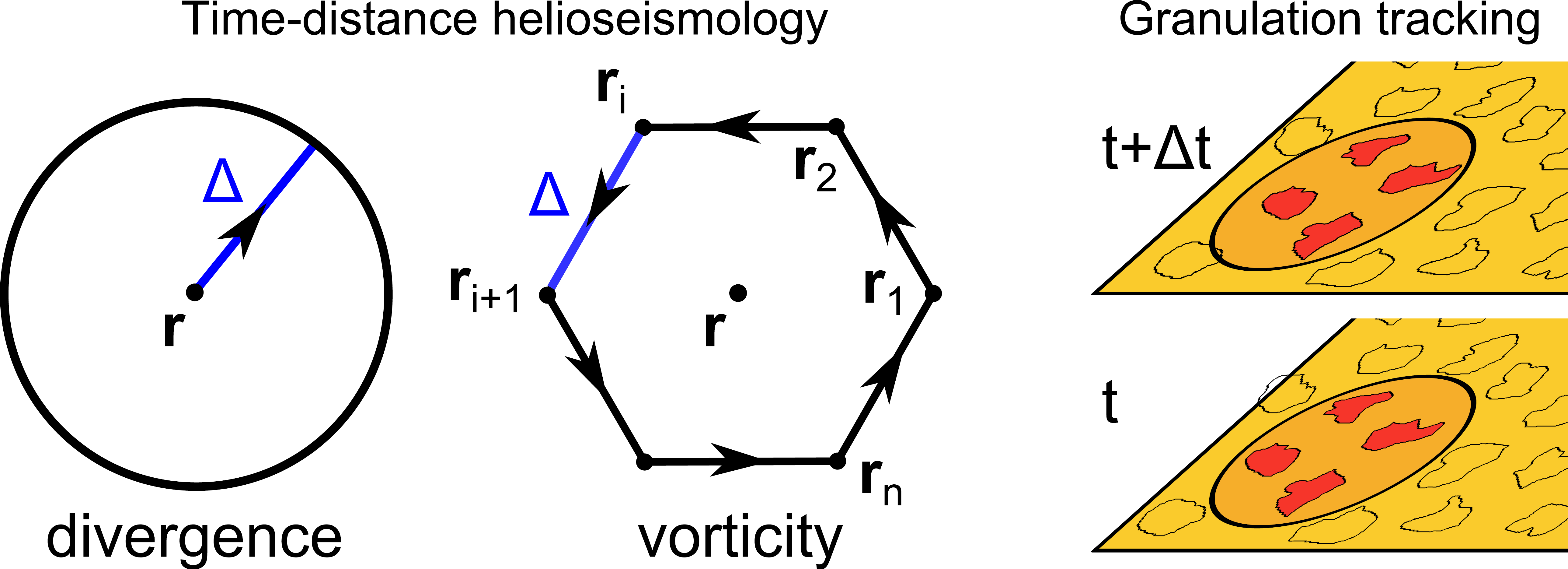}
\caption{Measurement techniques in local helioseismology and local correlation tracking (LCT).
In time-distance helioseismology, travel times of seismic waves measured between a central point and a concentric annulus inform us about the horizontal divergence of the flow, while travel times measured along a closed contour are related to the vertical vorticity of the flow. The motion of granules observed in time series of intensity images inform us about  horizontal flows at the solar surface. }
\label{fig.methods}
\end{figure}

\begin{figure}
\label{fig.divcurl}
\centering
\hspace*{\fill}
\begin{minipage}[t]{0.48\textwidth}
\includegraphics[width=\linewidth]{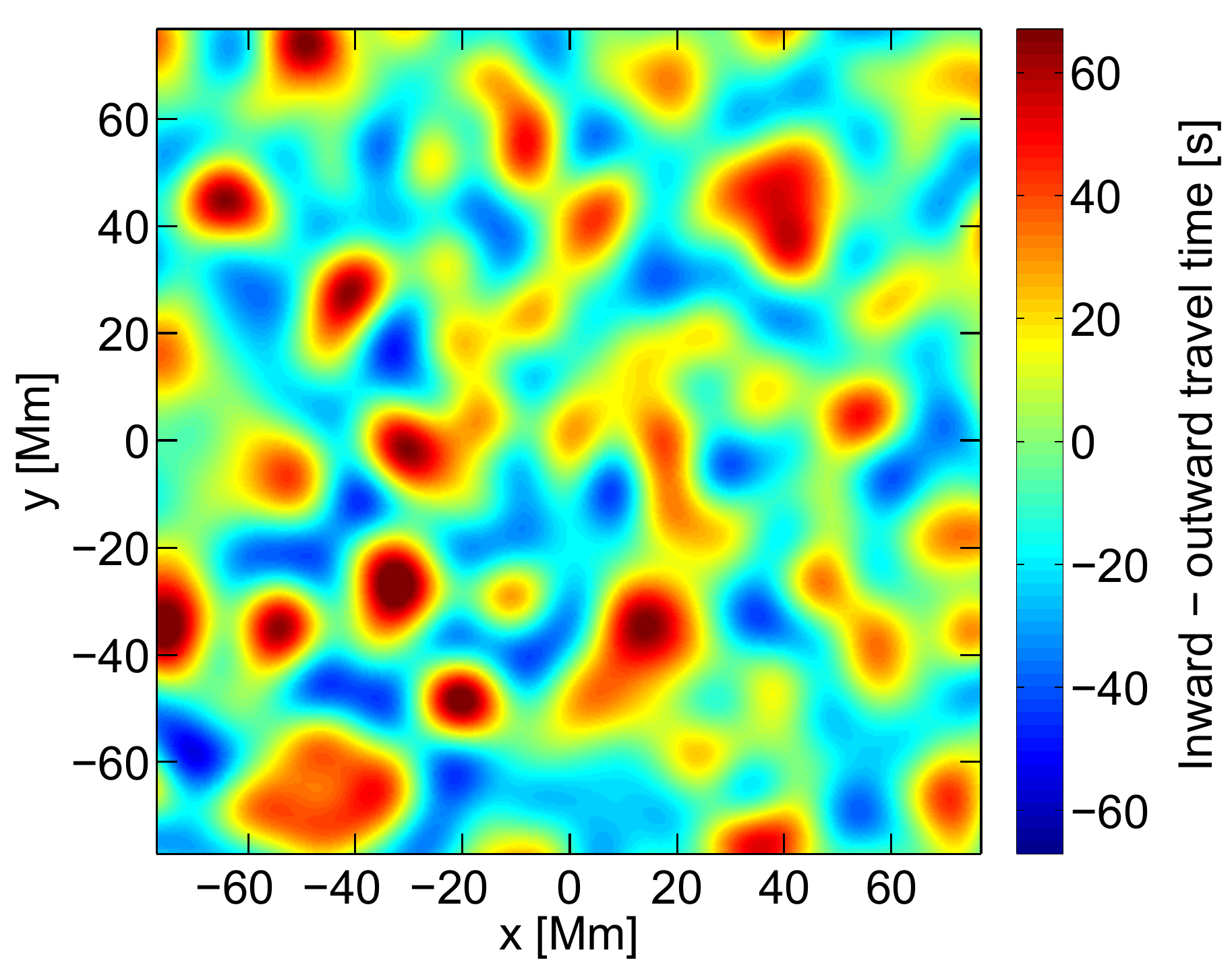}
\end{minipage}
\hspace*{\fill}
\begin{minipage}[t]{0.48\textwidth}
\includegraphics[width=\linewidth]{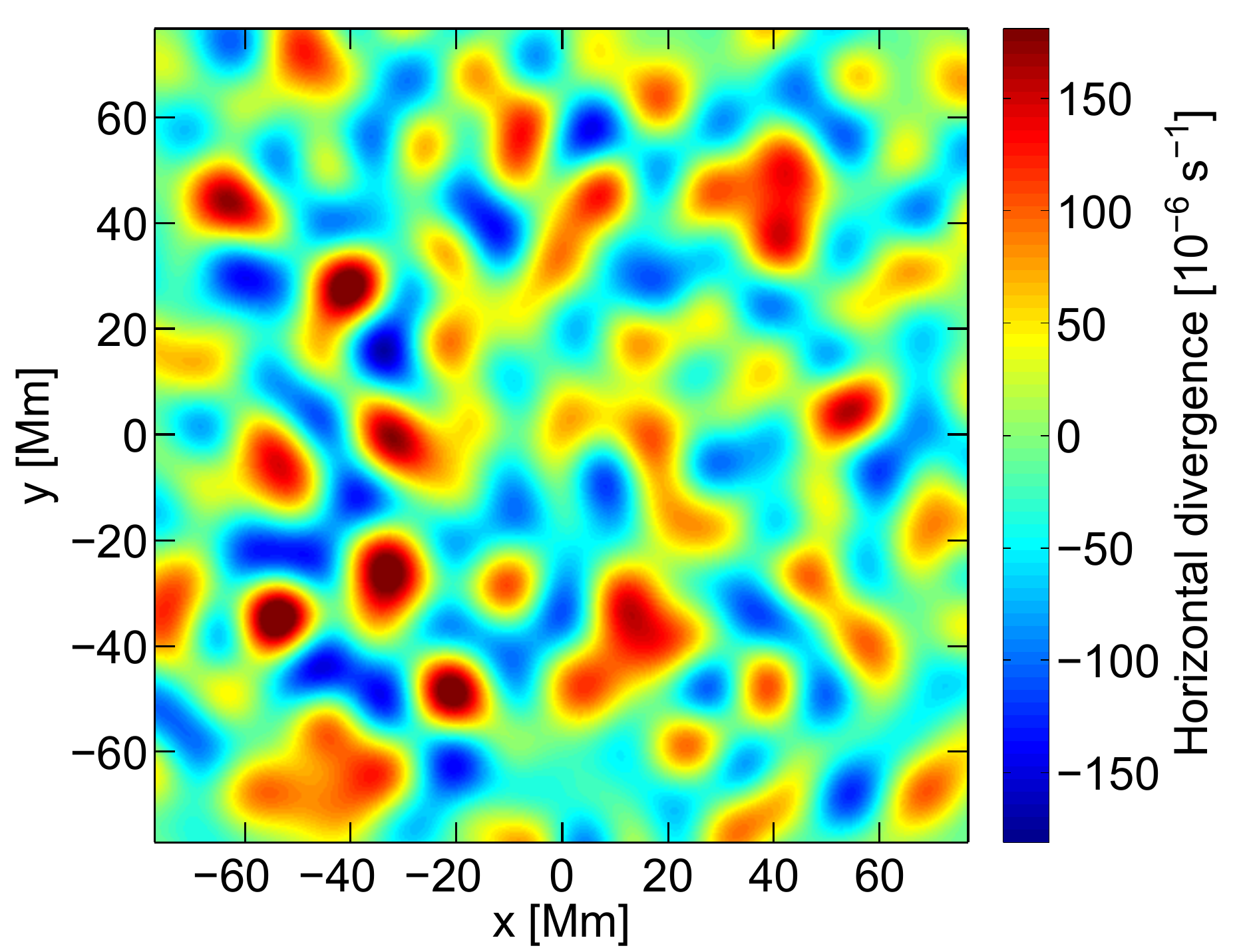}
\end{minipage}
\hspace*{\fill}
\begin{minipage}[t]{0.48\textwidth}
\includegraphics[width=\linewidth]{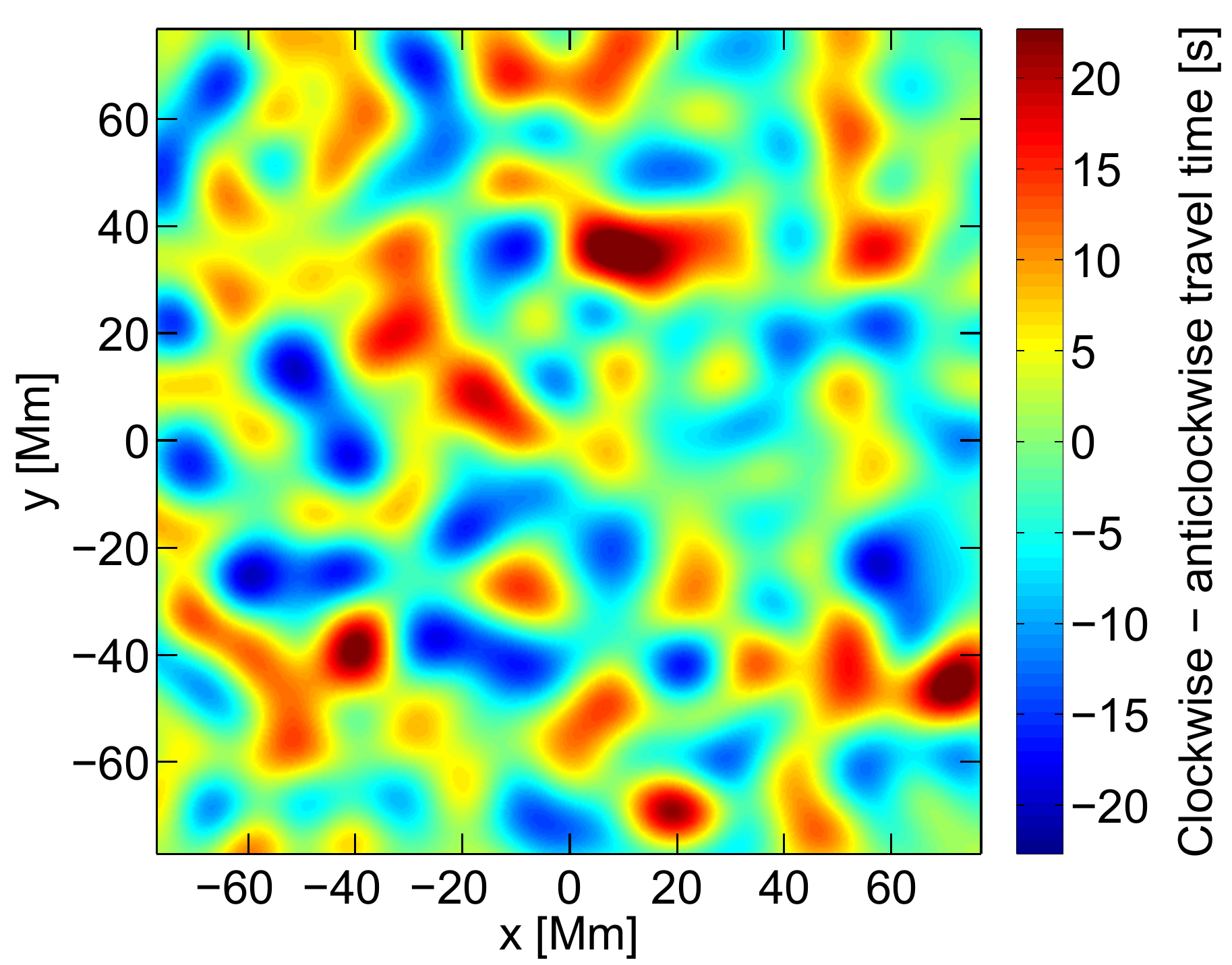}
\end{minipage}
\hspace*{\fill}
\begin{minipage}[t]{0.48\textwidth}
\includegraphics[width=\linewidth]{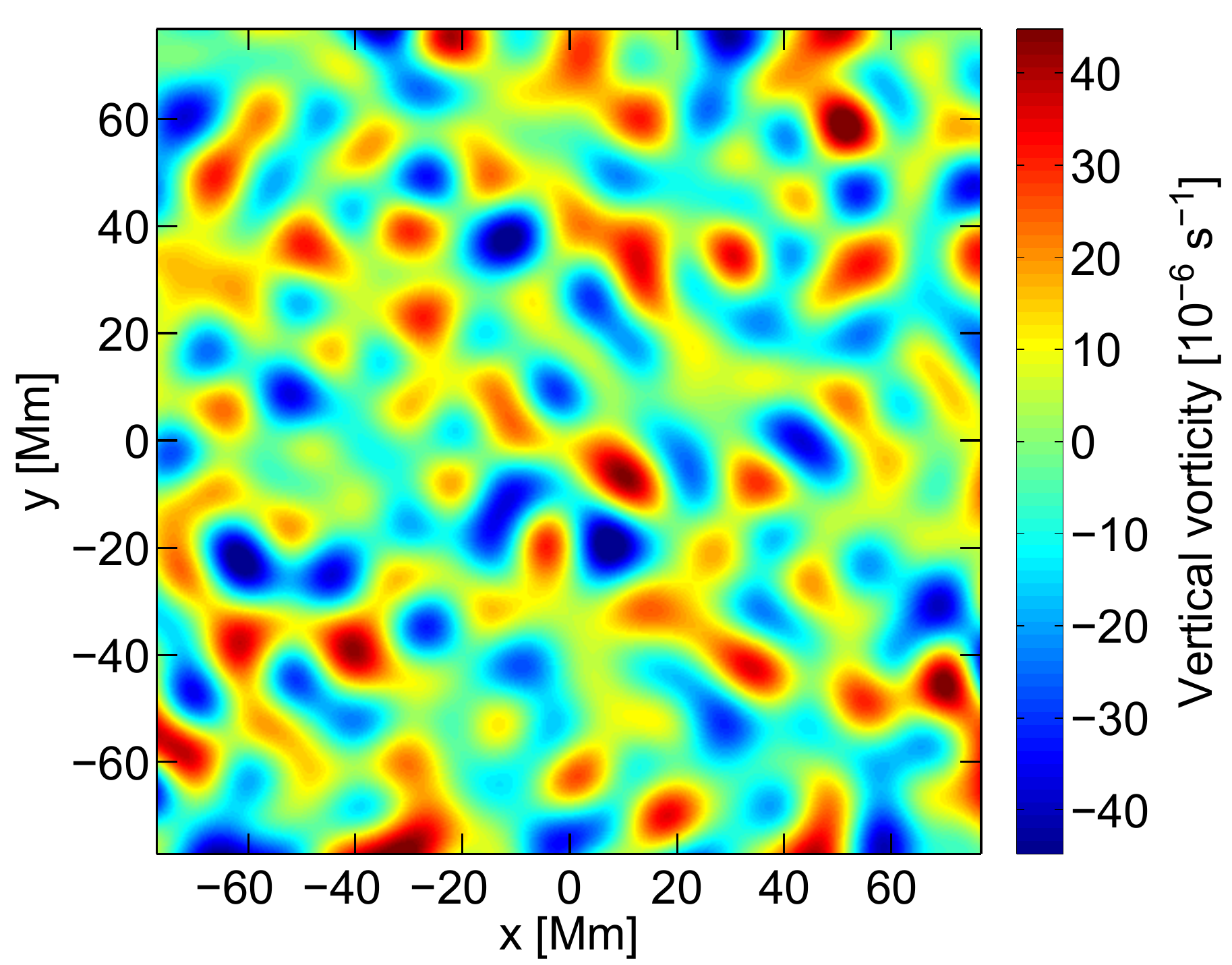}
\end{minipage}
\hspace*{\fill}
\caption{Horizontal divergence (top row) and radial vorticity (bottom row) of solar flows using HMI observations. The helioseismology (left panels, top 2~Mm) and LCT (right panels, surface) results are in good agreement: the correlation coefficients between the seismology and LCT maps are $0.93$ for the divergence and $0.54$ for the vorticity. The observation duration is eight hours and a low-pass filter  was applied ($l<300$). From \citet{Langfellner2014}. }
\end{figure}

Local helioseismology gives access to several useful quantities that describe turbulent convection in the near-surface layers: horizontal flow velocities, horizontal divergence, and two-point velocity correlations. 
\citet{Langfellner2014} measured the radial vorticity by computing the wave travel time around a closed contour. These quantities can help us characterize the dynamics and topology of rotating convection in the upper convection zone (from the surface down to a few Mm). For example, measurements of the radial vorticity provide a means to test  models of the influence of rotation on convection \citep[\eg][]{egorov2004}.  Away from the equator, \citet{Langfellner2014} observe a clear correlation between radial vorticity and horizontal divergence. The latitudinal dependence of this correlation, proportional to $\Omega \sin\lambda$, where $\Omega$ is the solar rotation rate at latitude $\lambda$, is consistent with the action of the Coriolis force on convective flows. 

In addition to local helioseismology, surface flow velocities can be measured with the local correlation tracking (LCT) of granules. The idea is to use granules as tracers of larger-scale flows. The LCT technique is very useful as it provides an independent measurement of horizontal  convective velocities at the solar surface. The method involves cross-correlating small patches containing a few granules in consecutive images measured in intensity (see Figure~\ref{fig.methods}). The time lag between consecutive images (typically 45~s) is much less than the lifetime of granules (10~min), thereby avoiding contamination from granulation evolution. Comparison of line-of-sight projected LCT velocity maps with Doppler and local helioseismology maps shows excellent agreement \citep{duvall00, derosa00}.
The spatial resolution of local helioseismology cannot be better than half the wavelength of seismic waves ($> 5$~Mm), while the resolution of the  LCT technique of the order of few granules ($>2$~Mm).
Both methods are able to probe spherical harmonic degrees less than about 300.

\subsection{Influence of rotation on convection}\label{rotation}

\begin{figure}
\label{fig.vortices}
\centering
\hspace*{\fill}
\begin{minipage}[t]{0.32\textwidth}
\includegraphics[width=\linewidth]{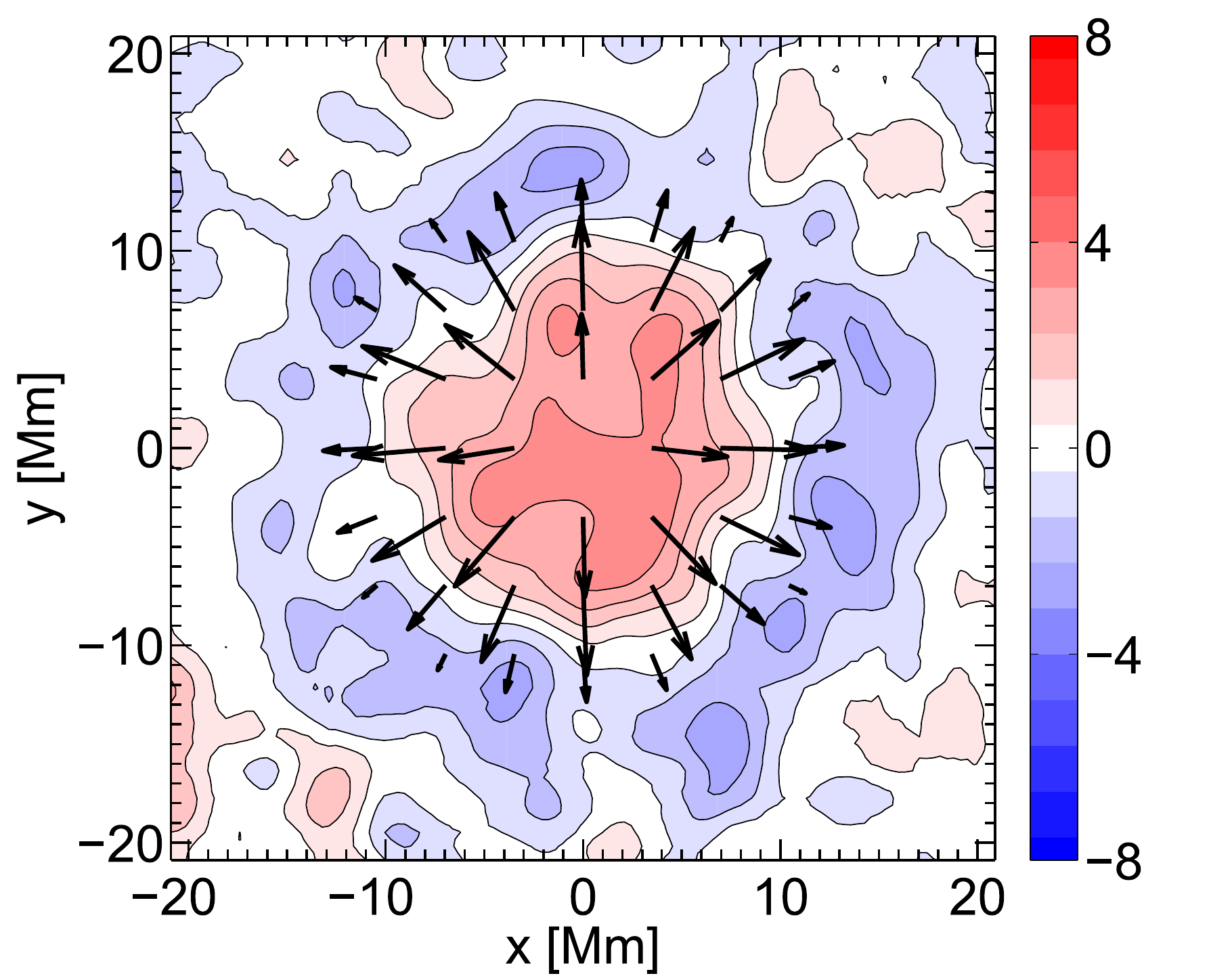}
\end{minipage}
\hspace*{\fill}
\begin{minipage}[t]{0.32\textwidth}
\includegraphics[width=\linewidth]{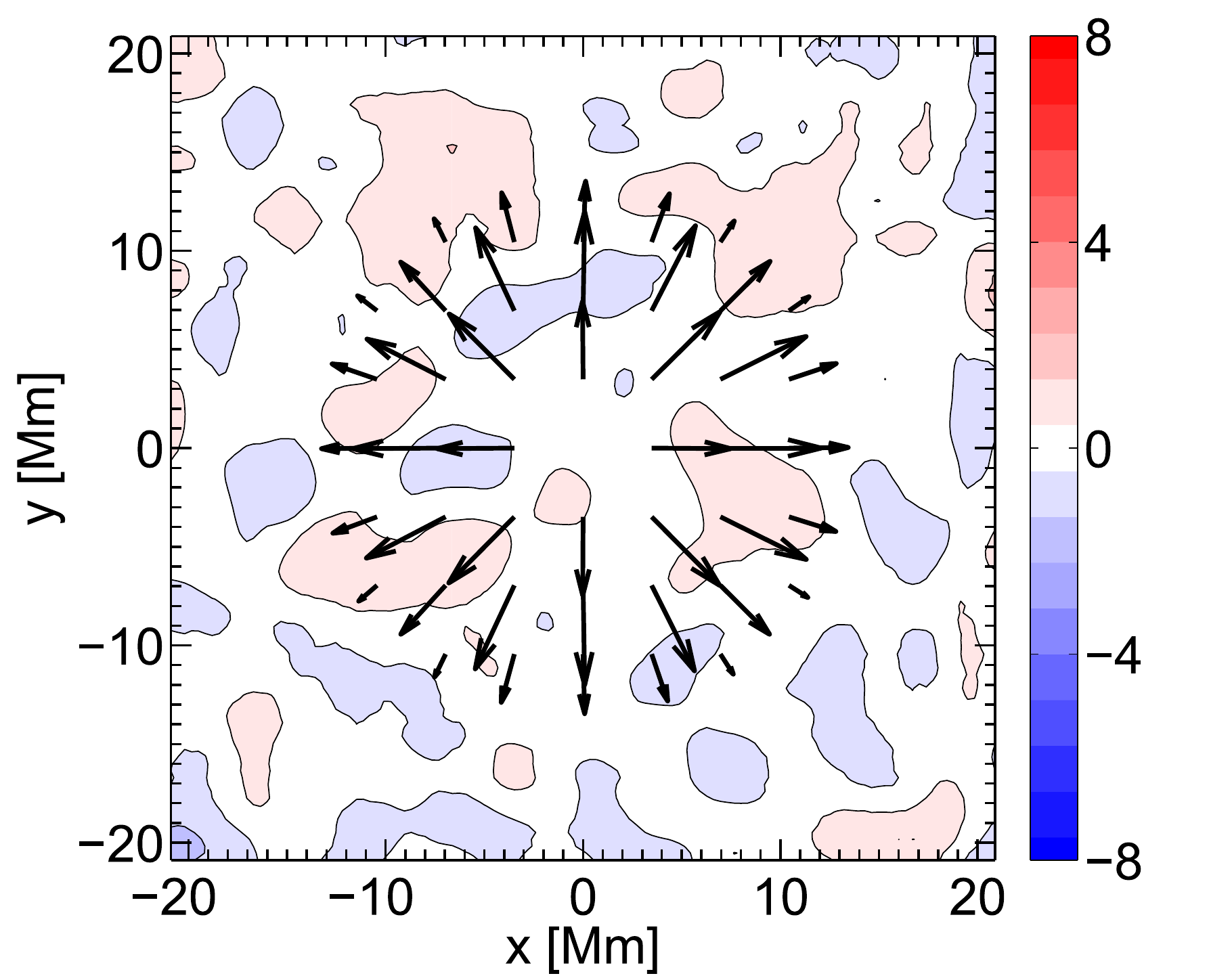}
\end{minipage}
\hspace*{\fill}
\begin{minipage}[t]{0.32\textwidth}
\includegraphics[width=\linewidth]{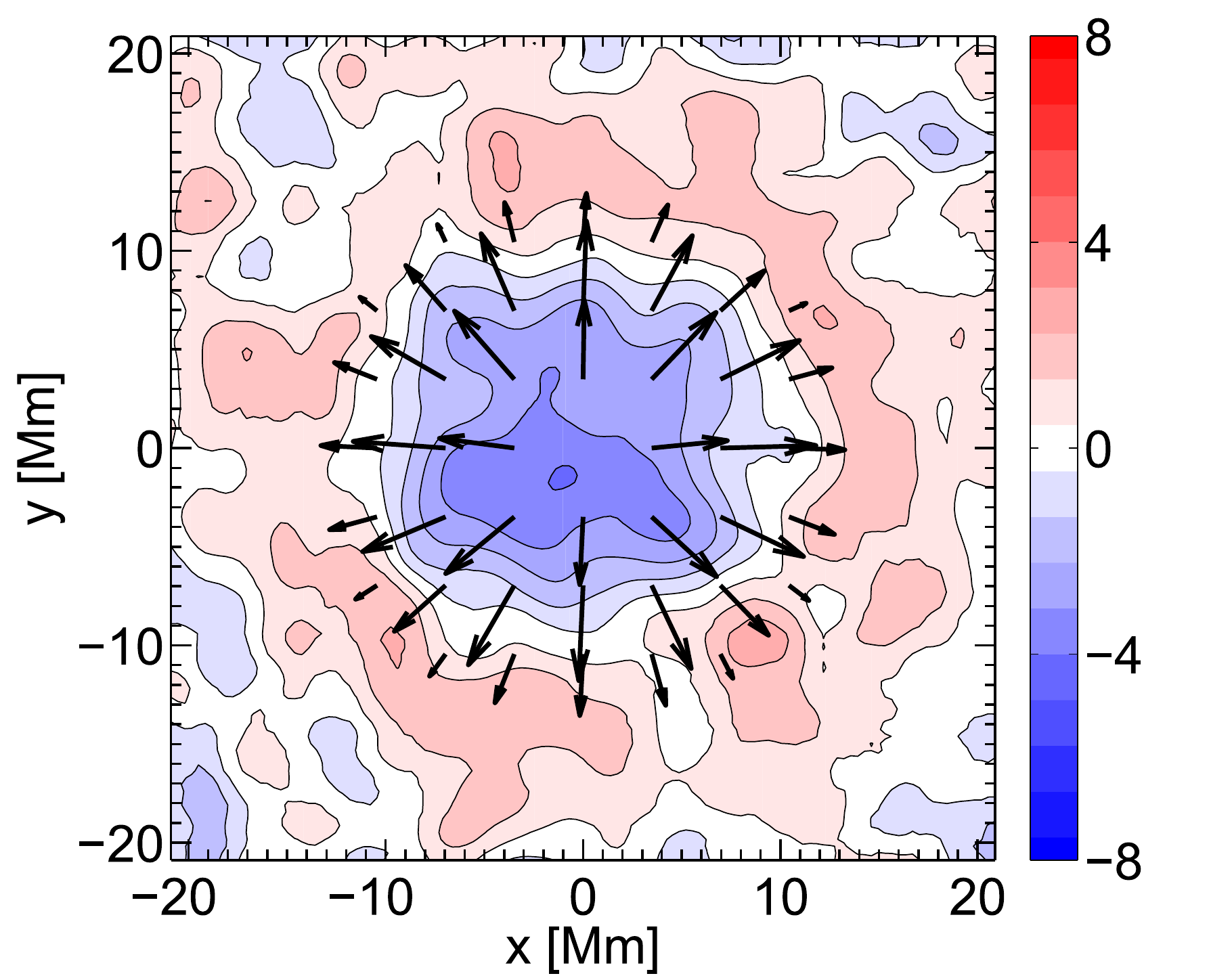}
\end{minipage}
\hspace*{\fill}
\begin{minipage}[t]{0.32\textwidth}
\includegraphics[width=\linewidth]{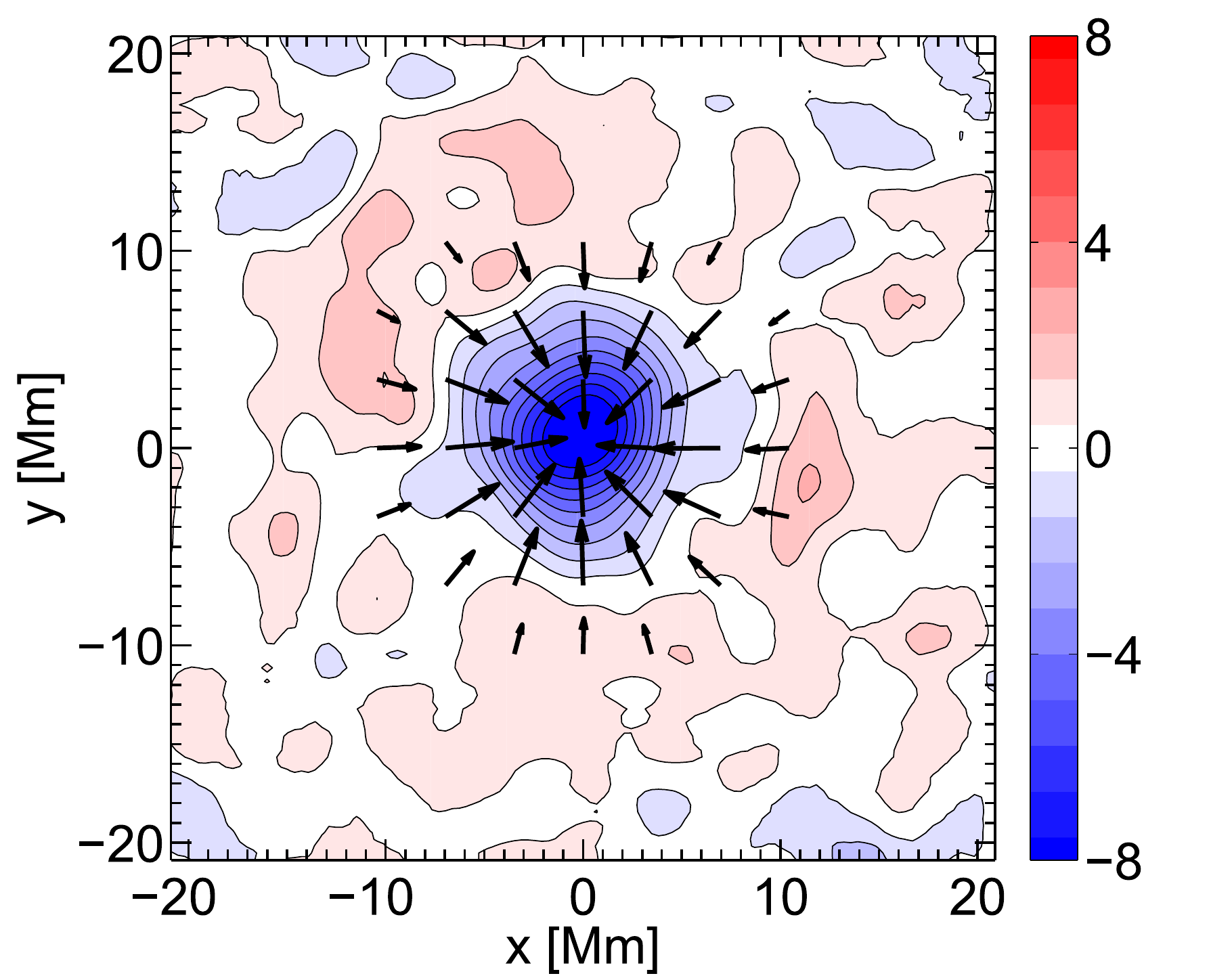}
\end{minipage}
\hspace*{\fill}
\begin{minipage}[t]{0.32\textwidth}
\includegraphics[width=\linewidth]{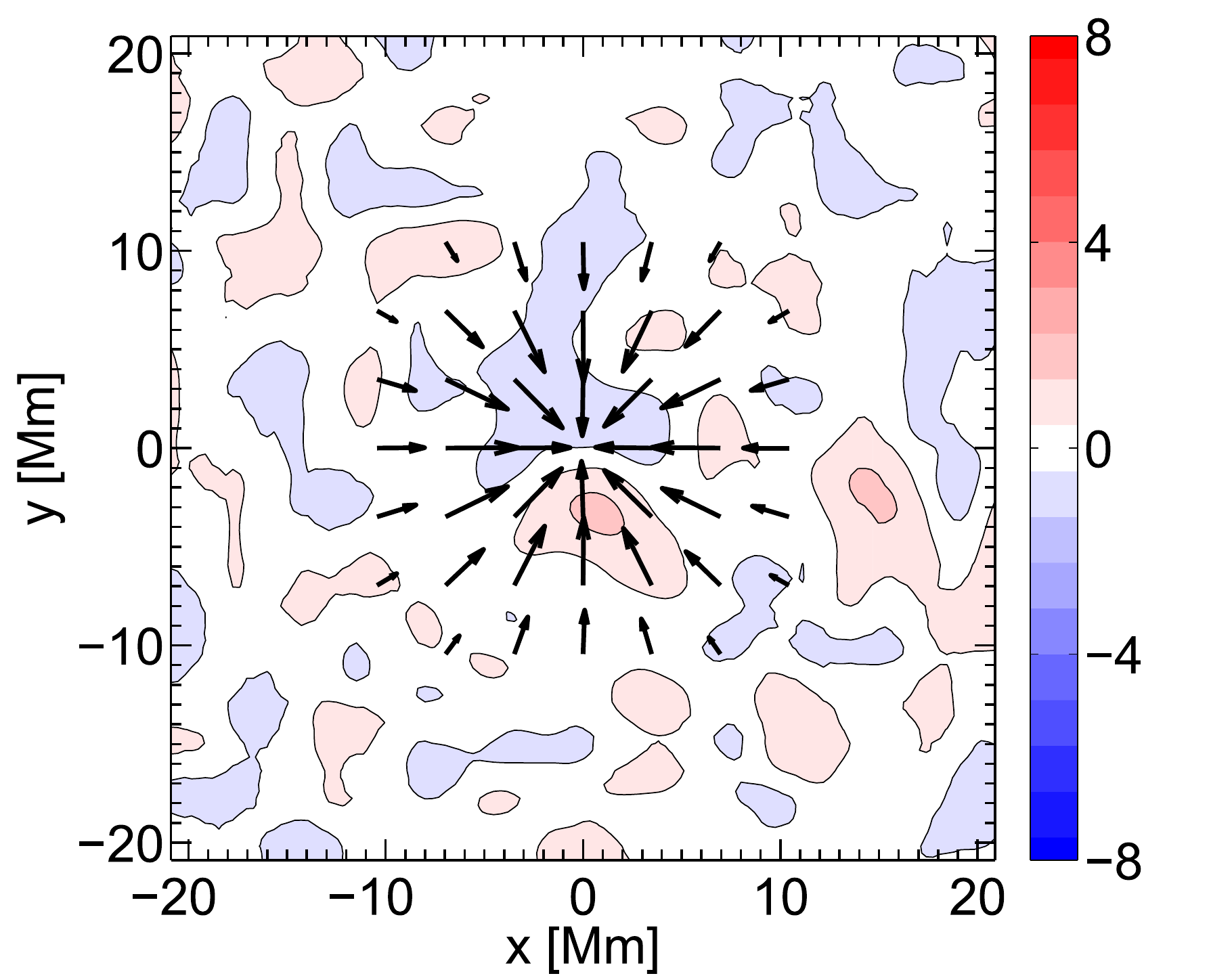}
\end{minipage}
\hspace*{\fill}
\begin{minipage}[t]{0.32\textwidth}
\includegraphics[width=\linewidth]{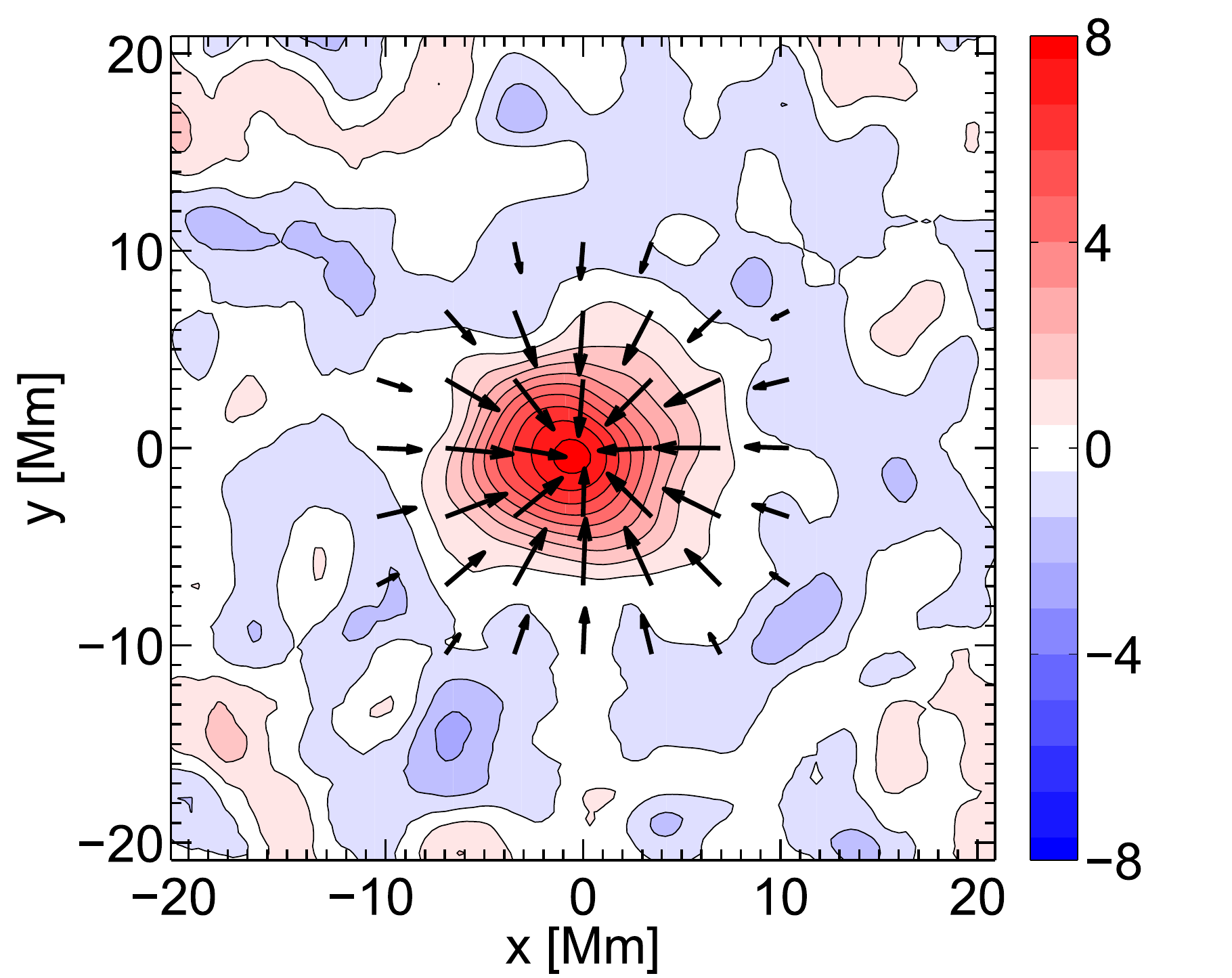}
\end{minipage}
\hspace*{\fill}
\caption{Vertical vorticity of the average supergranule in units of $10^{-6}$~s$^{-1}$
at latitudes $-40^\circ$, $0^\circ$, and $+40^\circ$ (from left to right) from local correlation tracking analysis. The colors refer to the radial component of vorticity, the arrows to the horizontal flow components.  The top row is for outflows, the bottom row for inflows.
Adapted from \citet{LAN15}. }
\end{figure}

The effects of rotation on convection depend on the Rayleigh and Coriolis numbers. 
The highest Rayleigh numbers at which rotating convection has been studied in the laboratory are $10^{11}<Ra<4.3\times10^{15}$ \citep{BNS}. The main result of this study is that modest rotation at high Rayleigh numbers has little effect on heat transport; if anything, rotation lowers the Nusselt number modestly. The vortical structure of the flows can however be affected quite significantly.

Convection in the Sun is influenced by rotation in a complex way. Large effects of rotation might have resulted in standard Taylor-Proudman convective columns parallel to the rotation axis. In contrast, a rather small effect is observed.  At smaller scales, the effects of rotation on convection become subtle as the Coriolis number becomes less than unity.  The supergranulation flows are associated with a 
 net kinetic helicity, $H_{\rm kin} = (\bnabla\wedge\bfv)\cdot\bfv$, and  the horizontal divergence is anti-correlated  with the radial vorticity in the northern hemisphere   \citep{Gizon2003a, gizon2010,Langfellner2014}. The correlation vanishes at the equator and switches sign in the southern hemisphere.

The positions of thousands of supergranules (in a latitude strip) can be read from the divergence maps using an image-segmentation algorithm. Averaging over outflow (or inflow) centers helps in lowering the noise level and in spatially resolving vorticity. Observations are shown in Figure 8 for the LCT technique. In the north, outflows regions are associated with clockwise motion, while inflows are associated with anti-clockwise motion. The azimuthal velocity in the vortices is on the order of $15$~m/s, i.e. 20 times smaller than the horizontal outflows. The patch of radial vorticity at the center of supergranules is more extended than in converging regions, where we observe a sharp localized peak in vorticity (width $\sim 5$~Mm).

\section{Future challenges}
\label{future}

\subsection{Differential rotation and Reynolds stresses}
The outer convective envelope rotates differentially as depicted in Figure~\ref{diffrot}. The inference of differential rotation in the convective envelope by helioseismology was remarkable because it revealed a profile that departs significantly from Taylor-Proudman balance (co-axial cylindrical iso-rotation).
In the limit of weak Reynolds stresses, it is possible to invoke thermal-wind balance for a rotating spherical fluid, which balances the latitudinal entropy gradient and the resulting rotation profile as
\begin{equation}
\frac{\partial\Omega^2}{\partial z} = \frac{g}{C_p  \, r\cos\theta}\frac{\partial\langle S\rangle}{\partial\theta},\label{balance}
\end{equation}
where $\theta$ is co-latitude, $r$ is radius, $z=r\cos\theta$ is the distance along the rotation axis, $\Omega = \Omega(r,\theta)$ represents the rotational shear in the convective envelope, $g$ is gravity, $\langle S\rangle = \langle S\rangle(r,\theta)$ is the temporally and longitudinally averaged entropy and $C_p$ is the specific heat at constant pressure. For fluid spheres for which the entropy is a constant, we recover the standard Taylor-Proudman balance, in which the rotation rate $\Omega$ is purely a function of the distance from the axis and contours of constant rotation would be parallel to the rotation axis. However, it is seen from Figure~\ref{diffrot} that the observed rotation rate is constant along conical contours, i.e. $\partial\Omega/\partial z \neq 0$, suggesting the prevalence of a latitudinal entropy gradient.

Equation~(\ref{balance}) by itself is unable to fully constrain the rotation rate since it allows for arbitrary terms that have no dependence on $z$ to be added to $\Omega^2$. The determining factor is the Reynolds-stress forcing, known in atmospheric-physics parlance by {\it gyroscopic pumping} \citep{miesch12}. Correlations between convective velocities,
\begin{equation}
R_{ij}  = \langle v'_i\, v'_j\rangle,
\label{restress}
\end{equation} 
yield the Reynolds-stress tensor ${\cal R}_{ij} = \rho R_{ij}$,
where $v'_i$ are the velocity fluctuations ($\langle v'_i\rangle=0$) and the angular brackets indicate longitudinal averaging. Measuring the Reynolds-stress distribution in the Sun is thus a major challenge and one with important consequences. The weak motion of large-scale convection (see section 3.3) already suggests that we may expect some unknown issues to arise.

\begin{figure}
\centering
\includegraphics[width=0.8\linewidth, trim= 80 360 70 80, clip]{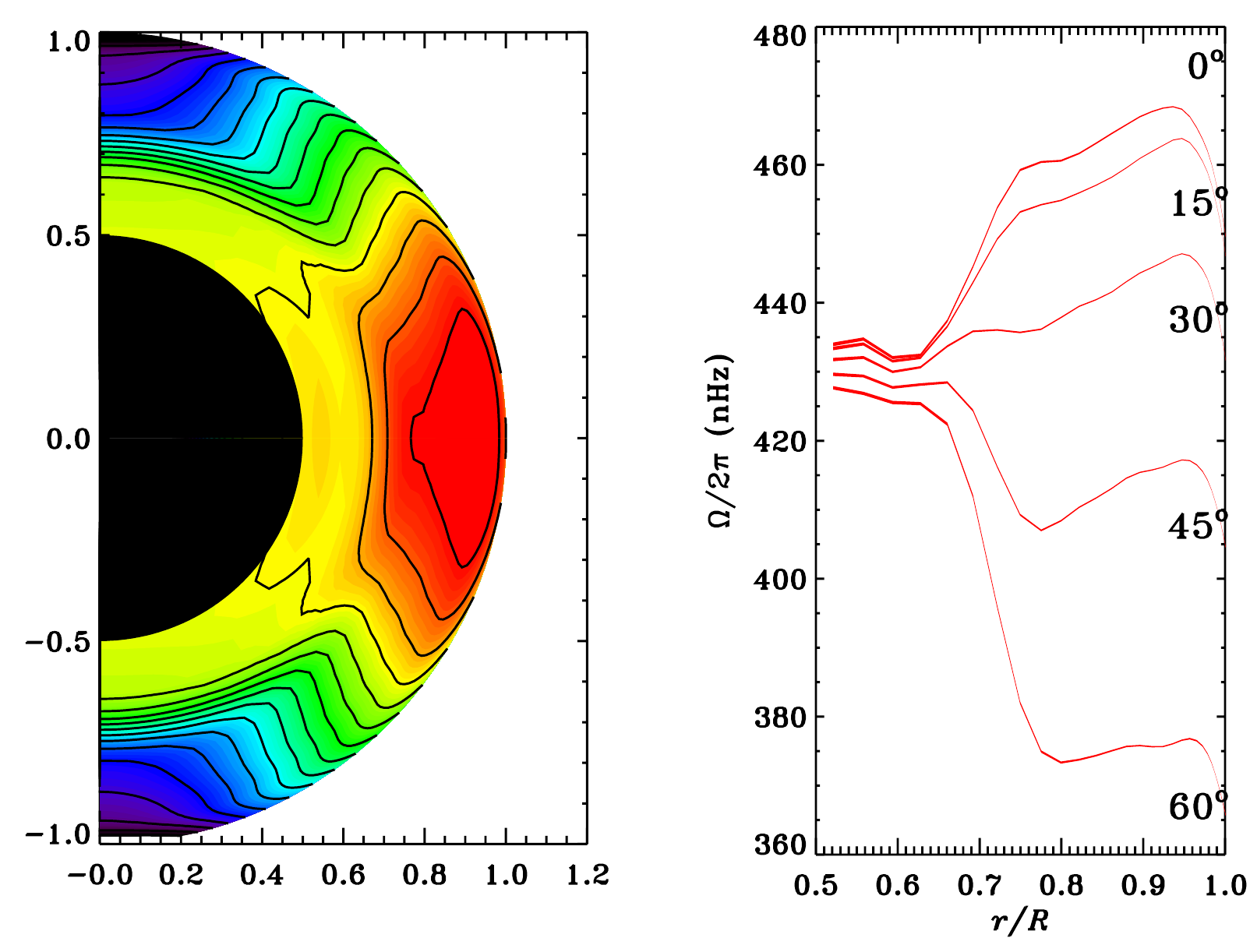}
\caption{
(Left) Contour plot of solar interior rotation obtained from 19 years of
data from the global network of ground-based observatories GONG. 
Black contours are at 10 nHz intervals. (Right) Rotational frequency as a
function of fractional radius at selected latitudes. The thickness of
the curves represents the formal (random) error. Note that due to
resolution issues the tachocline (the shear layer at the
bottom of the convection zone) is much thinner than shown here.
Based on work by \citet{howe09}. Courtesy of Rachel Howe.}
\label{diffrot}
\end{figure}

Models of differential rotation were developed e.g., by \citet{kitchatinov05} and  \citet{ruediger2014}, whereby angular momentum transport is due to small-scale rotating turbulence. In such mean-field models, $R_{\theta\phi}$ transports angular momentum toward the equator as required to produce solar-like differential rotation. 
The model predicts that $R_{\theta \phi} $ is the sum of a viscous term \citep{kitchatinov1994} and 
the so-called $\Lambda$ effect \citep[anisotropy of turbulent convection caused by rotation, see ][]{kitchatinov93}, with a latitudinal dependence of the form $R_{\theta \phi} \approx C \, \cos\theta \sin^2\theta$, where $\theta$ is co-latitude and $C$ depends on the relevant Coriolis number. 
\citet{hathaway2013} measured giant-cell velocity correlations at the surface  with $C_{\rm GC} \approx  50$~m$^2$\,s$^{-2}$, while seismic observations of supergranulation indicate a proportionality constant  $C_{\rm SG} \approx - 10^3$~m$^2$\,s$^{-2}$ (see Figure~\ref{fig.Rxy}a).  
Although the relative effect of rotation on supergranulation is  less than for giant cells (Coriolis number smaller), supergranulation velocities have larger amplitudes and thus measurable anisotropic Reynolds stresses. In the case of the (deep) giant cells, the $\Lambda$ effect (positive $C$)  dominates over the viscous term (negative $C$) and transports angular momentum towards the equator.  Closer to the surface, however, the viscous term dominates in the supergranulation layer (Figure~\ref{fig.Rxy}b). Observations and model are in remarkable agreement.

\begin{figure}[t]
\centering
\begin{minipage}{.49\textwidth}
  \centering
    \includegraphics[width=\linewidth]{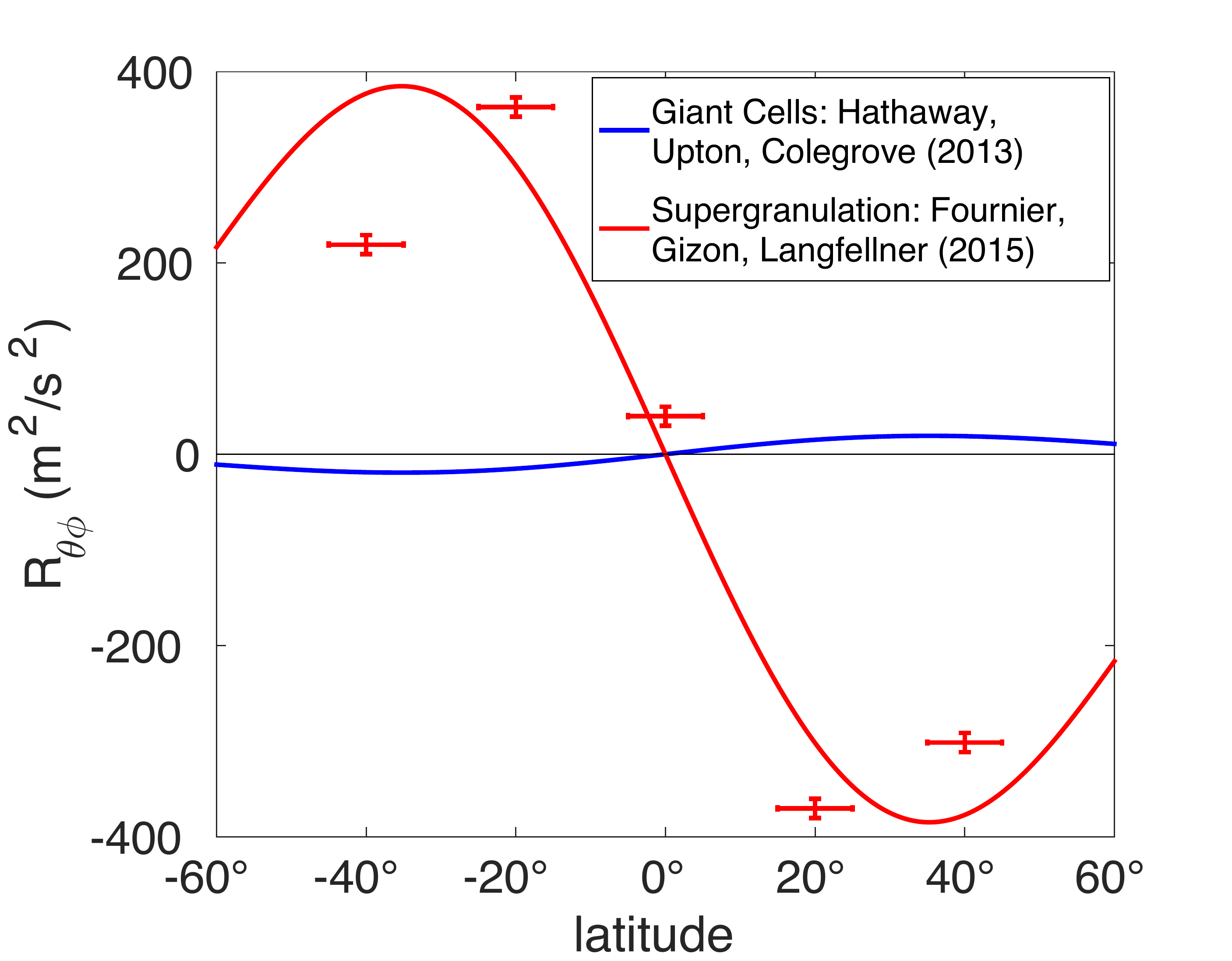}
\end{minipage}
\begin{minipage}{.39\textwidth}
  \centering
   \includegraphics[width=\linewidth]{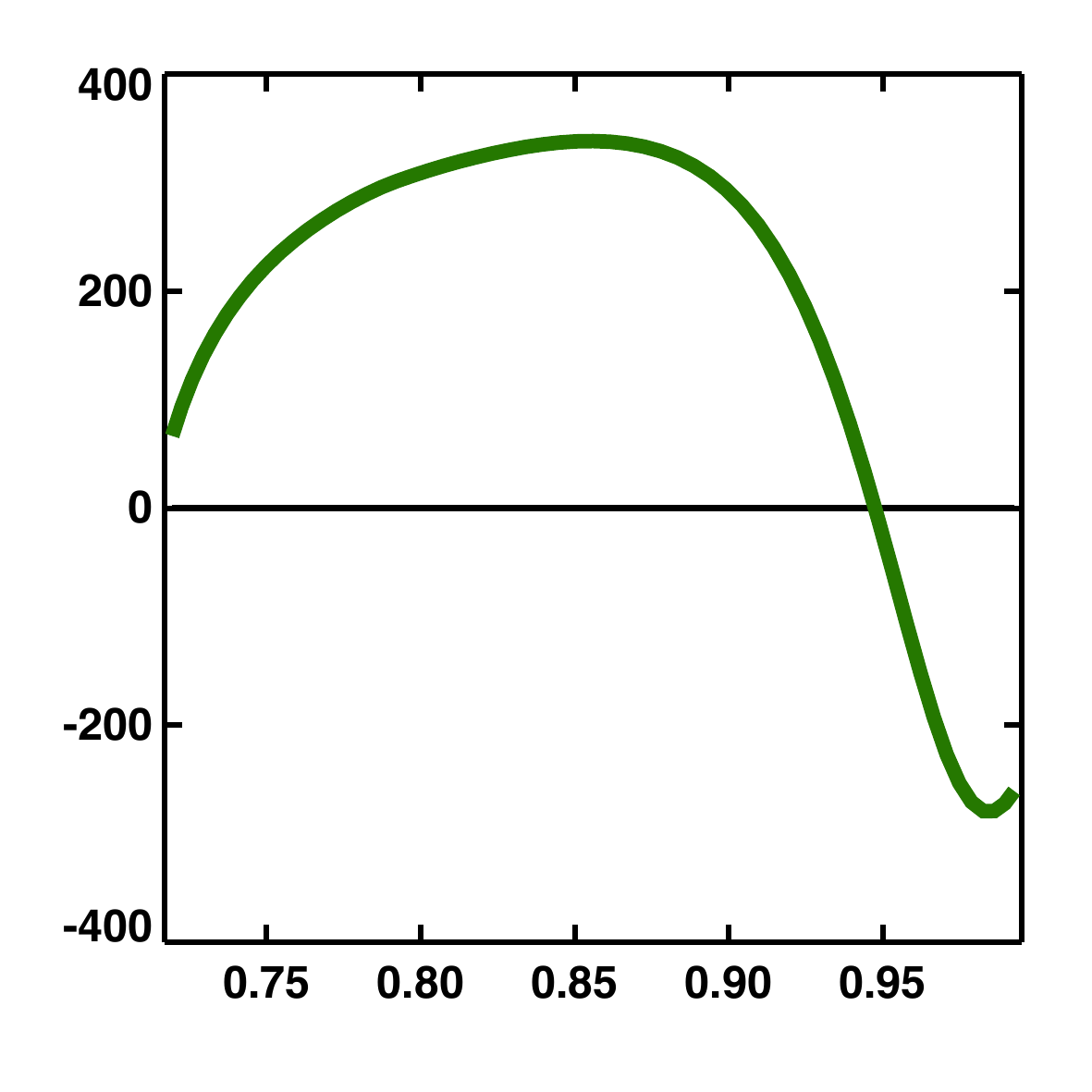}
\end{minipage}
\caption{({\it Left}) Observed  horizontal average $R_{\theta\phi}=\langle v'_\theta v'_\phi  \rangle$ of co-latitudinal (southward) and longitudinal (prograde) fluctuating velocities  as a function of solar latitude. 
The blue curve is for giant cells \citep[local correlation tracking of supergranulation,][]{hathaway2013} and the data points and the red curve are for supergranulation (helioseismology, courtesy of Fournier, Gizon \& Langfellner). The  functional form $\cos \theta \sin^2\theta$ is assumed.
({\it Right})   $R_{\theta\phi}$ at latitude $30^\circ$ north as a function of fractional radius from a mean field model \citep[cf.][]{ruediger2014}. Units of m$^2$\,s$^{-2}$ are the same as in the left panel. 
Courtesy of Manfred K\"uker.
} \label{fig.Rxy}
\end{figure}

\begin{figure}
\centering
\hspace*{\fill}
\begin{minipage}[t]{0.29\textwidth}
\includegraphics[width=\linewidth]{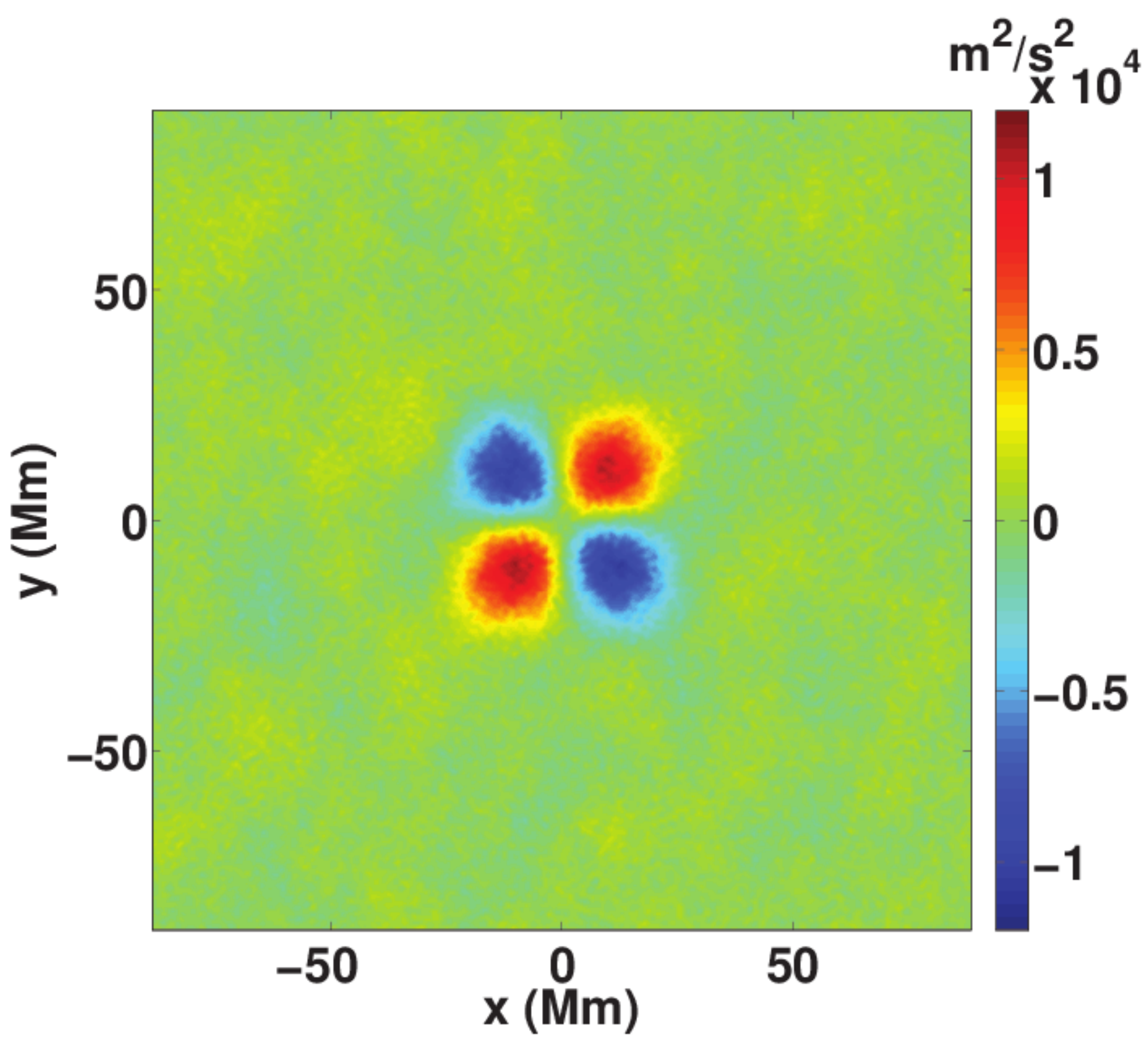}
\end{minipage}
\hspace*{\fill}
\begin{minipage}[t]{0.29\textwidth}
\includegraphics[width=\linewidth]{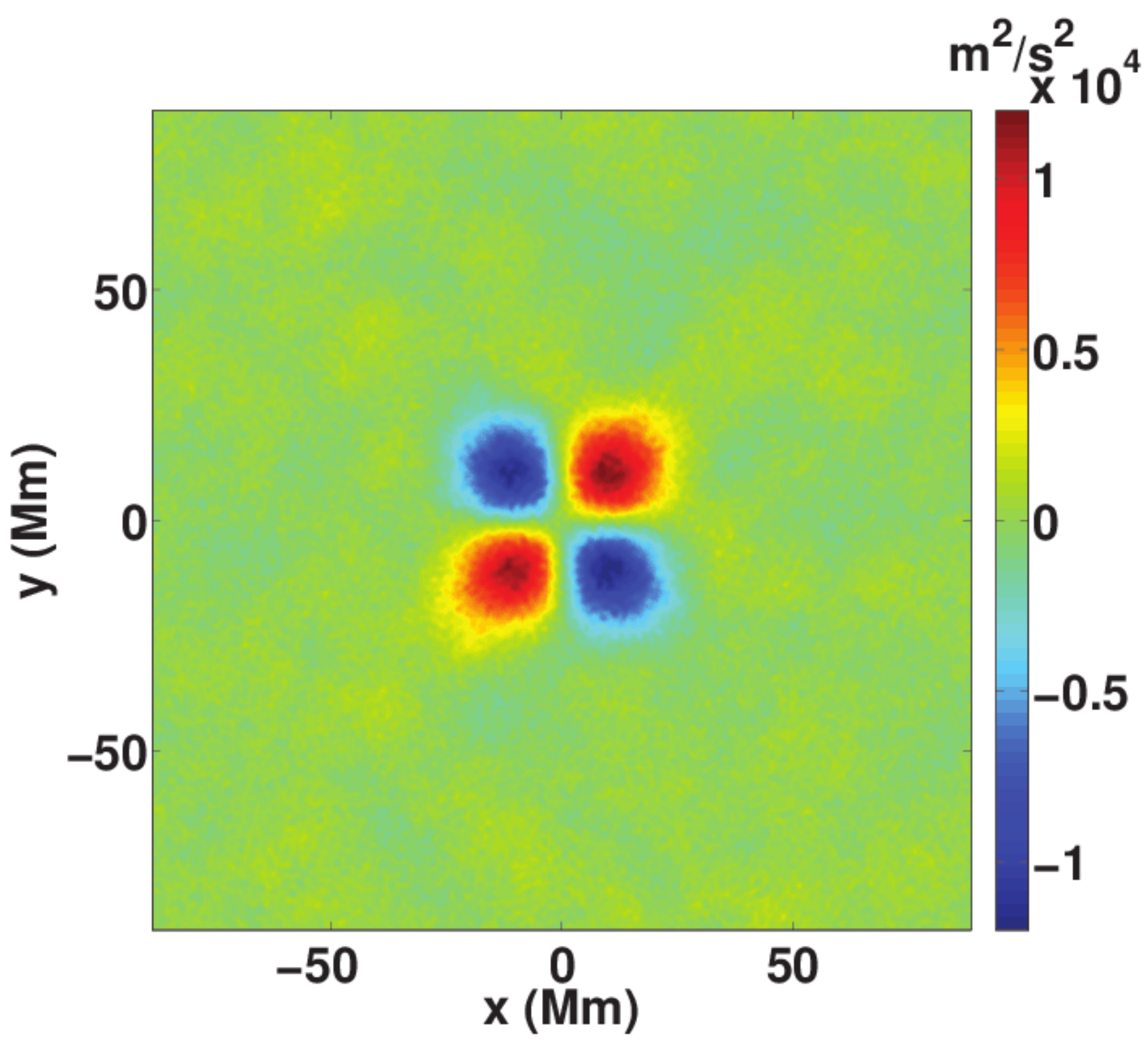}
\end{minipage}
\hspace*{\fill}
\begin{minipage}[t]{0.29\textwidth}
\includegraphics[width=\linewidth]{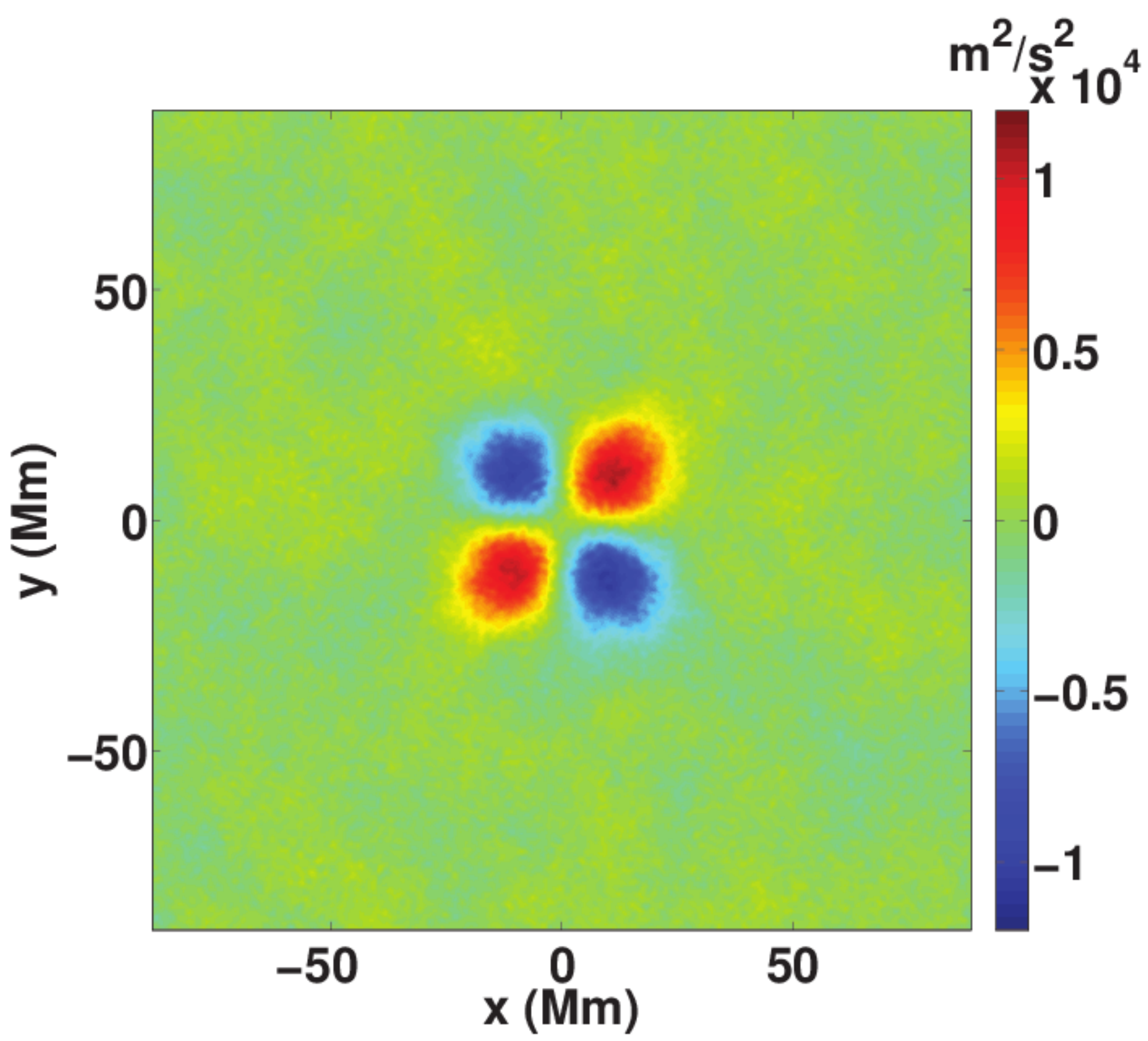}
\end{minipage}
\hspace*{\fill}
\begin{minipage}[t]{0.29\textwidth}
\includegraphics[width=\linewidth]{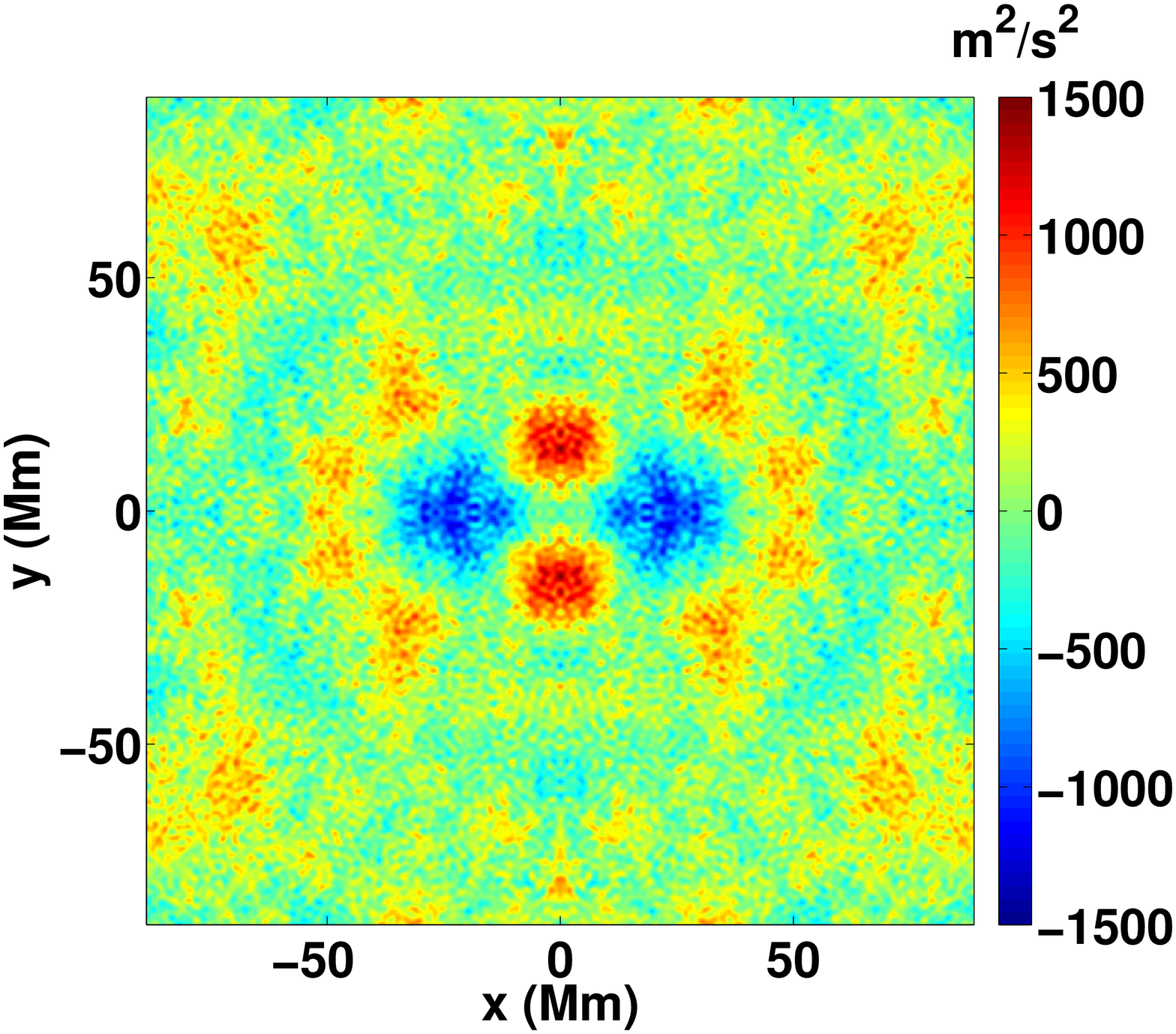}
\end{minipage}
\hspace*{\fill}
\begin{minipage}[t]{0.29\textwidth}
\includegraphics[width=\linewidth]{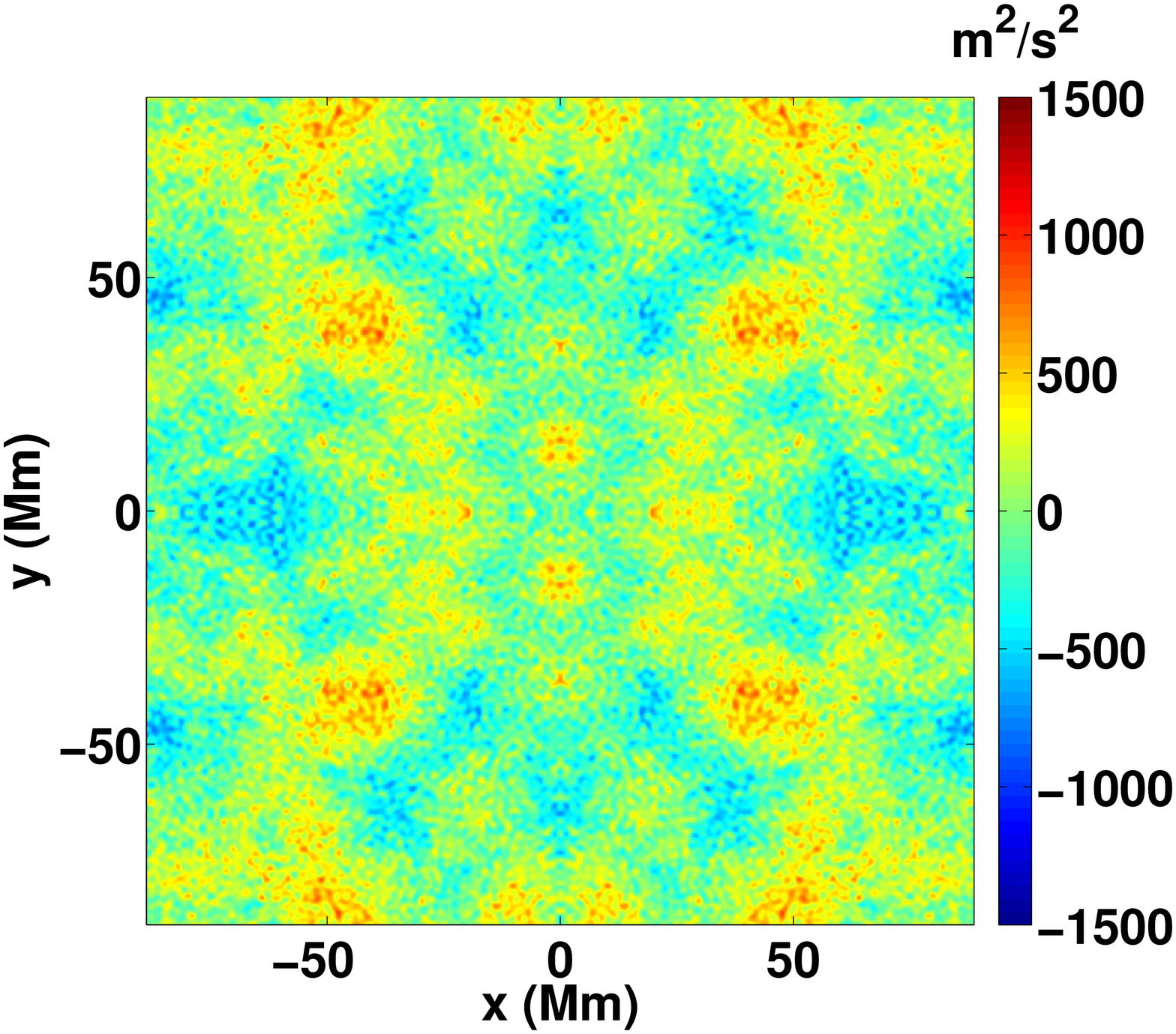}
\end{minipage}
\hspace*{\fill}
\begin{minipage}[t]{0.29\textwidth}
\includegraphics[width=\linewidth]{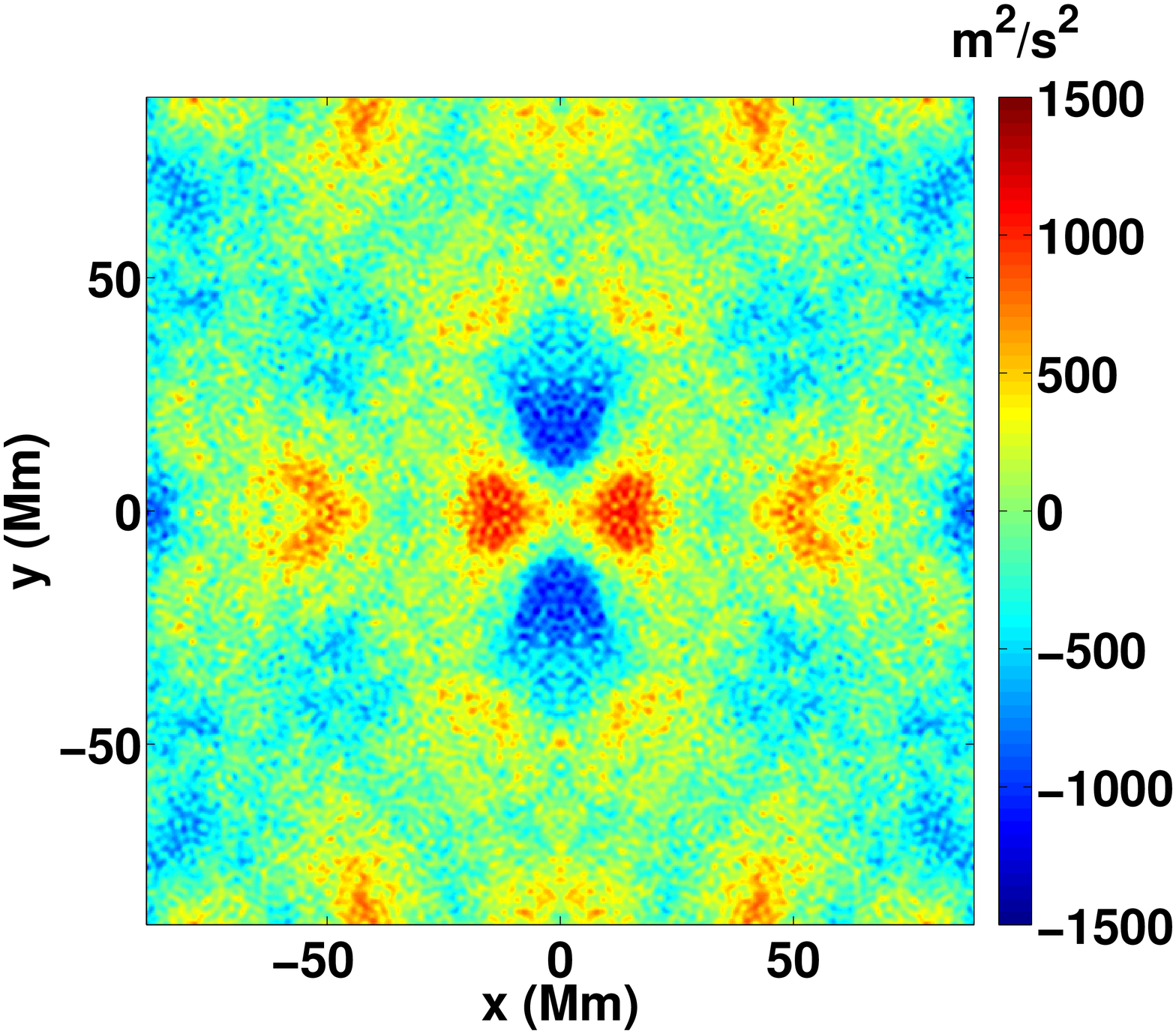}
\end{minipage}
\caption{  ({\it Top panels})  Two-point velocity correlation function, $R_{xy}(\bx)= \langle v_x'(\bx') v_y'(\bx'+\bx) \rangle$, as a function of horizontal lag $\bx=(x,y)$, from local helioseismology and for latitudes $-40^\circ$, $0^\circ$, and $40^\circ$ (from {\it left} to {\it right}). The coordinates $x$ and $y$ point in the directions of longitude and latitude respectively. The main features are due to non-rotating convection.  ({\it Bottom panels}) Rotation-induced component of $R_{xy}(\bx)$ at latitudes $-40^\circ$, $0^\circ$, and $40^\circ$ (from {\it left} to {\it right}), obtained by taking the mirror-symmetric component of $R_{xy}$.  Based on the work by Fournier, Gizon, and Langfellner.
 }
\label{fig.velcorr2}
\end{figure}

The model of \citet{kitchatinov93} not only predicts the components of the Reynolds-stress tensor, but more generally the two-point velocity correlations
\begin{equation}
R_{ij}(\bx) = \langle   v'_i (\bx')  v'_j  (\bx'+\bx) \rangle ,
\end{equation}
where $\bx$ is the spatial lag and angular brackets denote an average over $\bx'$ in a patch of the Sun around latitude $\lambda$. 
The perturbation to $R_{\theta\phi}(\bx)$  introduced by rotation, denoted by $\delta R_{\theta\phi}(\bx)$, is expected to be mirror symmetric and to depend on the latitude through the local Coriolis number. In order to check the predictions of the model, measurements of the two-point correlation were obtained using time-distance helioseismology (Figure~\ref{fig.velcorr2}). We find that $\delta R_{xy}$ has the correct symmetries, has different signs in the northern  and southern hemispheres, and  vanishes at the equator. 

One of the interesting dynamical properties of the upper layers of the Sun is the existence of a strong radial gradient of rotation \citep[\eg][]{Schou2003}. In this layer, which occupies about 5\% of the solar convection zone, the rotational velocity increases with growing depth at the rate $\sim 2$~m/s per Mm. The latest measurements by \citet{Barekat2014} using {HMI observations} show that the logarithmic gradient of the rotation rate ($d\ln \Omega/d \ln r$) near the surface is close to $-1$ from the equator up to $60^\circ$ latitude. 

Both local heliseismology and LCT present serious possibilities for measuring $R_{ij}$ near the surface.  Future work aims at including a larger range of spatial scales and  probing deeper in the convection zone. The $r$-components of the Reynolds-stress tensor are more difficult to measure due to the fact that radial velocity fluctuations $v_r'$ are very small.

\subsection{Large-scale magnetism}
The Sun supports a global magnetic field that undergoes cyclic polarity reversal on a time-scale of roughly 11 years (in comparison, the geodynamo is erratic and reverses polarity on timescales of thousands of years). Elucidating the mechanism of the solar dynamo is a central goal of solar physics. A widely held description of the solar dynamo is the Babcock-Leighton mechanism \citep{dikpati99}, which relies on differential rotation to intensify magnetic fields buried at the base of the convection zone. Rotational shear would strengthen these fields till they become magnetically buoyant and rise to the surface, where they would ostensibly form sunspots and other large-scale magnetic complexes. Magnetic tubes are much like rubber bands and the amplification process would require angular momentum to be extracted from rotational shear. However, differential rotation is observed to be remarkably constant \citep[with small fluctuations, \eg][]{howe00}, and this would indicate that shear in angular momentum is being continuously replenished by convective Reynolds stresses, as is generally believed.

\subsection{Supergranulation}

Supergranulation is a prominently visible scale of convection on solar surface (Figure~\ref{supergran}) though the thermal contrast is weak; they are only slightly warmer, likely less than 3 K at their centers \citep[\eg][]{Meunier2007, rast09}. Consequently, the driving mechanism of supergranulation and its role in global solar dynamics is a matter of some debate \citep[see][for a review]{rieutord10}.

It is not known definitively whether supergranules are deep or shallow structures (connected to deep or shallow convection processes). Local helioseismology shows that they penetrate at least down to a depth of 5~Mm \citep[\eg][]{Sekii2007, jason08}. However, one cannot exclude the likelihood that supergranulation downdrafts remain coherent over much greater depths. The seismic inference of the internal structure of supergranulation provides an effective means of discriminating between theories. Thus, supergranulation has been a subject of central interest in helioseismology \citep[\eg][]{Kosovichev1997, duvall00,  
braun04, Svanda2011, dombroski13}.

\citet{duvall06} pioneered the statistical study of ensembles of supergranules, based on the assumption of horizontal homogeneity. Analyzing and averaging wave travel times over thousands of supergranules, \citet{birch06} obtained precise estimates for the travel times associated with the average supergranule. These travel times can ostensibly be used to create accurate models of supergranules. However, seismic inferences have been unable to arrive at a unified model that successfully explains observed travel times. A number of these efforts are at odds with each other, and analyses by \citet{duvall13}, \citet{svanda12} and \citet{duvall14} have suggested a previously unanticipated result, namely that supergranular flows peak at a depth of 2\,--\,3 Mm, reaching up to 800 ${\rm ms^{-1}}$ in horizontal flow speed. In contrast, \citet{birch06} and \citet{woodard07} infer supergranular flows that decrease in magnitude with depth. Much work, in terms of refining solutions to both the forward and inverse problems in local helioseismology, is needed to satisfactorily address these issues.

\citet{Duvall1980} showed that the supergranulation pattern rotates 3--4\% faster than the local plasma. This phenomenon is known as the superrotation of supergranules. The evolution of the supergranulation pattern was analyzed by \citet{gizon03} using long sequences (two to three months each year) of seismic maps of horizontal divergence of the flows (see space-time slices in Figure~\ref{fig.SGslices}). Fourier analysis of these data   revealed that the supergranulation pattern has wave-like properties with oscillation periods in the range of 6--9 days and excess power in the direction of rotation, in accordance with an independent analysis based on direct Doppler imaging \citep{Schou2003}.
\citet{gizon03} suggested that supergranulation is traveling-wave convection: the supergranulation pattern  can be modeled by a superposition of modes (with anisotropic power) that travel at a phase speed of about 65~m/s.

\begin{figure} \centering 
\begin{overpic}[width=\linewidth, trim=140 485 140 15, clip]{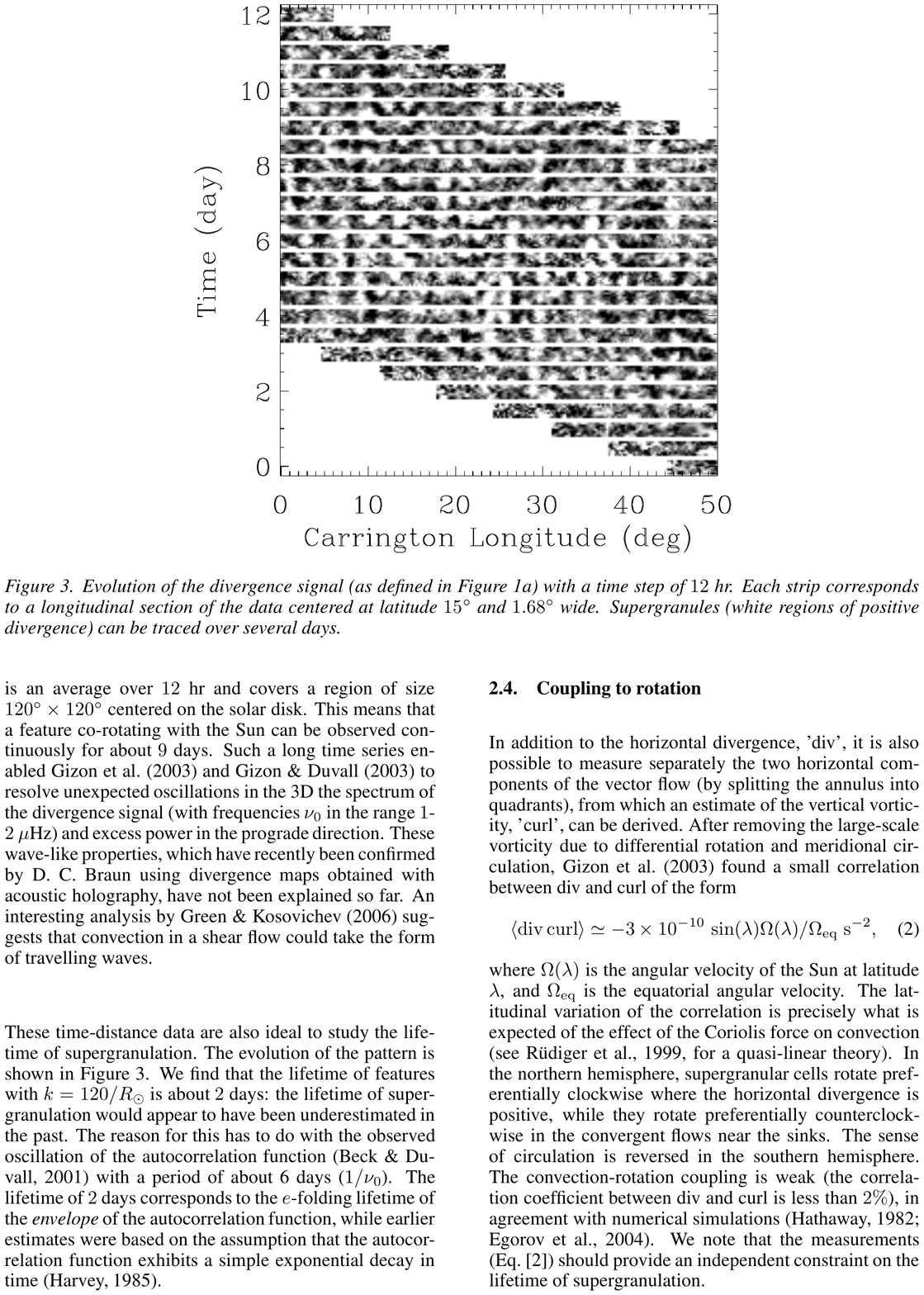} 
\linethickness{2pt}
\put(59.5,20){\color{red}\vector(0.08,1){2.5}}
\end{overpic}
\caption{Evolution of the supergranulation pattern (divergence of horizontal flows) versus  Carrington longitude and time. The Carrington longitude is defined in a frame that co-rotates with the Sun. Each strip corresponds to a longitudinal section of the data centered at latitude $15^\circ$ and $1.68^\circ$ wide. Supergranules (white regions of positive divergence) can be traced over several days.  Several supergranules are long lived and are seen to travel faster (red arrow) than the local rotation rate.  From \citet{Gizon2006}. } \label{fig.SGslices} \end{figure}

Since \citet{Chandrasekhar1961}, who investigated the linear stability of thermal convection, it is known that time-dependent oscillatory modes can develop in thermal convection in the presence of either rotation or the magnetic field.  Furthermore, the presence of shear significantly influences convectively unstable modes, which become traveling modes and propagate horizontally. \citet{Adam1977} studied a convectively unstable layer bounded below and above by stable layers and found that the phase speed of the traveling convection modes can cover the full range imposed by the shear flow. \citet{Busse2003, Busse2007} showed that rotating convection cells will in general exhibit a drift in the direction of rotation and generate a mean flow. We note that the solar near-surface shear flow ($\sim 70$ m/s) is comparable to the phase speed of the supergranular modes ($65$~m/s). Whether this is a coincidence or a physical effect should be investigated, possibly via numerical simulation.

\section{Summary points}
\begin{itemize}
\item Convection in the Sun operates in an extraordinary parameter regime, ${Re}\sim~10^{14}, {Ra}\sim10^{20}, {Pr}\sim10^{-6}$.
\item Outstanding challenges in understanding the Sun lie in estimating the Reynolds stresses and establishing the mechanism governing the emergence of large-scale, fluid circulations (such as differential rotation and meridional circulation) and the sustenance of its global magnetic field.
\item Traditional means for understanding these challenges have relied on mixing length theory and important numerical simulations (which, however, do not resolve the flow field even remotely well).
\item The Sun's interior is optically inaccessible but supports acoustic waves, well approximated by linear theory.
\item Seismic waves are advected by the underlying convecting medium, thereby shifting their travel times, which have been used to infer motions of the medium.
\item Helioseismology has been used to constrain the amplitudes of convective velocities at large scales ($\ell < 60$) to be two orders in magnitude lower than current theory and computation, throwing into question our understanding of thermal and angular momentum transport in the solar convection zone.
\item Seismology is also used to image intermediate scales of convection such as supergranulation ($\ell \approx 120$) and to measure the anisotropic Reynolds stresses that control the global dynamics of the solar convection zone.
\end{itemize}

\section{Future issues}
\begin{itemize}
\item A complete seismic analysis of convective velocities that addresses a range of depths, from the mid-convection zone to the near-surface layers while simultaneously accounting for sources of theoretical and observational errors is critical. Although a difficult problem, it will be a significant advance in placing seismic constraints on convection on a rigorous footing.
\item Numerical simulations of solar convection miss some critical ingredients since they are unable to accurately reproduce solar differential rotation; it is unlikely that one can push the simulations to the appropriate parameters, so one needs to have some independent and quantitative means for contrasting the numerical data. The path forward likely lies in the identification of important and relevant physical processes that control the dynamics of the convection zone and simulating these phenomena.
\item Laboratory experiments can lend significant insight towards appreciating the properties of highly stratified and non-Boussinesq convection, but studies of this sort are very sparse.
\end{itemize}

\acknowledgements
This work is supported in part  by the NYU-Abu Dhabi Center for Space Sciences.  SH acknowledges a collaboration with the Max Planck Institute for Solar System Research through a Max Planck Partner Group established at the Tata Institute of Fundamental Research, Mumbai.  LG acknowledges support from Deutsche Forschungsgemeinschaft SFB 963 ``Astrophysical Flow Instabilities and Turbulence''.  This work utilizes  HMI data that are courtesy of NASA/SDO and the HMI science team and are made available through the German Data Center for SDO, funded by the German Aerospace Center (DLR). The GONG program, managed by the National Solar Observatory,  is operated by AURA, Inc. under a cooperative agreement with the National Science Foundation; the data were acquired by instruments operated by the Big Bear Solar Observatory, High Altitude Observatory, Learmonth Solar Observatory, Udaipur Solar Observatory, Instituto de Astrof\'{\i}sica de Canarias, and Cerro Tololo Interamerican Observatory. We thank Aaron Birch, Damien Fournier, David Hathaway, Rachel Howe, Manfred K\"uker, Jan Langfellner, Mark Miesch, and Robert Stein for providing figures, as well as Robert Cameron, John Leibacher, and Olga Shishkina for useful discussions.

\section{Annotated references}
\begin{itemize}
\item {\bf \citet{duvall}} introduced the method of time-distance helioseismology, widely used to image three-dimensional structures and flows in the solar interior. 
\item {\bf \citet{hanasoge12_conv}} bounded solar large-scale convective velocities, constraining them to be two orders in magnitude smaller than theory.
\item {\bf \citet{hmi}} described the purpose and workings of the Helioseismic and Magnetic Imager (HMI)  onboard the Solar Dynamics Observatory (SDO),  launched by NASA in 2010.
\item {\bf \citet{miesch05}} gave an overview of numerical simulations of solar convection.
\item {\bf \citet{gizon2010}} summarized  methods and results of local helioseismology.
\item {\bf \citet{niemela00}} performed convection experiments using liquid Helium at extremely high Rayleigh numbers.
\end{itemize}

\section{Related Resources}
\begin{itemize}
\item HMI data are publicly available at http://hmi.stanford.edu and from the German Data Center at http://www2.mps.mpg.de/projects/seismo/GDC-SDO
\end{itemize}

\section{Sidebar}
{\bf Alongside Section 1.1}\\
Much like the Sun, the Earth also possesses a convection zone (albeit in a vastly different parameter regime), termed the mantle, where overturning convection is driven by heating from the core. Surface cooling results in the formation of a crust and at subduction zones (such as the Pacific ``ring of fire"), pieces of the crust break off and fall back into the interior. Terrestrial seismic imaging suggests that these pieces of crust (likened to the descending plumes in the Sun) reach the core-mantle boundary intact \citep[e.g.][]{masters91}. Based on this picture, it is tempting to speculate that descending cool plumes formed at the solar surface maintain their identity on their journey to the base of the convection zone. This suggests that convection is not well mixed in the solar interior, indicating small convective velocities, in line with seismic and related inferences. It must, however, be recognized that the parameter regimes are starkly different; the Reynolds number is very small within Earth's mantle and the Prandtl number is on the order of $10^2$ (and dramatically varying with the depth).


\section{Acronyms and definitions} 
\begin{itemize}
\item ASH: Anelastic Spherical Harmonic code, that simulates global convection in the Sun in the anelastic limit.
\item LCT: Local Correlation Tracking is a technique that uses granules to trace larger-scale flows. It provides an independent measurement of horizontal  convective velocities at the solar surface.
\item GONG: Global Oscillation Network Group, a network of ground-based telescopes around the world that observes the Sun continuously.
\item HMI: Helioseismic and Magnetic Imager, the helioseismology instrument onboard SDO
\item SDO: Solar Dynamics Observatory, a NASA satellite in geosynchronous orbit launched in 2010. 
\item Rayleigh Number: $Ra$ is the ratio of buoyancy to viscous forces multiplied by the ratio of viscous to thermal diffusivities. Beyond a critical value of the Rayleigh number, convection sets in; below this value, conduction is the primary mechanism of thermal transport. The mathematical definition can be more complicated for stratified fluids.
\item Reynolds Number: is the ratio of inertial to viscous force, $Re = U L / \nu$, where $U$ is the characteristic convective velocity, $L$ the characteristic length scale and $\nu$ the kinematic viscosity.
\item The Coriolis number is the ratio of convective turnover time $(L/U)$ to rotation period $2\pi/\Omega$, where $\Omega$ is the rotation rate, i.e. $Co = L\Omega / (2\pi U)$. It is inversely related to the Rossby Number, which is the ratio of inertial to Coriolis force, $Ro = U/(2 L\Omega)$, by the relation $Co \sim 1/(4\pi Ro)$. 
\item Prandtl Number: the ratio of viscous to thermal diffusivities, $Pr = \nu/\alpha$,.
\end{itemize}

\bibliographystyle{ar-style1.bst}

\bibliography{seismology-convection}
\end{document}